\documentclass[fleqn,usenatbib]{aa}

\usepackage[varg]{txfonts}
\usepackage{graphicx}	
\usepackage{amsmath}	
\usepackage{amssymb}	
\usepackage{ae,aecompl}

\begin{document}

\title{Modelling the Canes Venatici~I dwarf spheroidal galaxy}

\author{
  D. R. Matus Carillo\inst{1} \and
  M. Fellhauer\inst{1} \and
  A. G. Alarcon Jara\inst{1,2} \and
  C. A. Aravena\inst{1} \and
  F. Urrutia Zapata\inst{1}}
\institute{
  Departamento de Astronom\'ia, Universidad de Concepcion,
  Casilla 160-C, 3340 Concepci\'on, Chile \and
  Observatories of the Carnegie Institution of Washington, 813
  Santa Barbara St., Pasadena, CA 91101, USA}

\date{Received XX.XX.XXXX / Accepted XX.XX.XXXX}

\abstract{
  The aim of this work is to find a progenitor for Canes Venatici~I
  (CVn~I), under the assumption that it is a dark matter free object
  that is undergoing tidal disruption.  With a simple point mass
  integrator, we searched for an orbit for this galaxy using its
  current position, position angle, and radial velocity in the sky as
  constraints.  The orbit that gives the best results has the pair of
  proper motions $\mu_\alpha = -0.099\,\mathrm{mas\ yr^{-1}}$ and
  $\mu_\delta = -0.147\,\mathrm{mas\ yr^{-1}}$, that is an apogalactic
  distance of $242.79$~kpc and a perigalactic distance of $20.01$~kpc.   
  Using a dark matter free progenitor that undergoes tidal disruption,
  the best-fitting model matches the final mass, surface brightness,
  effective radius, and velocity dispersion of CVn~I simultaneously.
  This model has an initial Plummer mass of $2.47 \times
  10^7\,\mathrm{M_\odot}$ and a Plummer radius of $653$~pc, producing a
  remnant after $10$~Gyr with a final mass of $2.45 \times
  10^5\,\mathrm{M}_\odot$, a central surface brightness of
  $26.9$~mag\,arcsec$^{-2}$, an effective radius of $545.7$~pc, and 
  a velocity dispersion with the value $7.58$~km\,s$^{-1}$.
  Furthermore, it is matching the position angle and ellipticity of
  the projected object in the sky.} 


\titlerunning{Modelling CVn~I}
\authorrunning{Matus Carillo et al.}

\maketitle

\section{Introduction}
\label{sec:intro}

Dwarf spheroidal (dSph) galaxies are one of the most abundant objects
in the Universe \citep{mateo_1998, metz_2007b}.  As the name implies,
dSph are small galaxies, with a low stellar content.  They have low
amounts of gas or even lack it completely, have no star formation, and
have a population of very old ( $> 10$~Gyr) and metal poor ($-3 <
[\mathrm{Fe/H}] < 0$) stars \citep{mateo_1998, mcconnachie_2012}. 
According to the $\Lambda$ Cold Dark Matter model, massive galaxies,
like the Milky Way (MW), should be surrounded by large numbers of dark
matter (DM) dominated satellites \citep{simon_2007}.  Their small
number of stars makes them very faint, with absolute magnitudes
between $-13 \leq M_{V} \leq -7$ magnitudes \citep{mateo_1998,
  belokurov_2007}, and hard to detect \citep{tollerud_2008}; $\sim 50$
of them are identified as satellites of the MW
\citep{mcconnachie_2012, koposov_2015, newton_2017}.  Despite their
low luminous mass, dSphs have high velocity dispersions
\citep{mateo_1998, simon_2007}.  Assuming virial equilibrium,
spherical symmetry, and isotropic velocity dispersion, the dynamical
mass of these galaxies are of the order of $10^7$--$10^8\
\mathrm{M}_\odot$, within their half-light radius.  Since the
dynamical mass is much higher than the luminous mass (stars), the 
general consensus is that they contain a high amount of DM.  In fact
they are the objects with the highest concentration of DM in the
Universe, reaching  mass-to-light ratios (M/Ls) of $\sim 100$ or even
$\sim 1000$ \citep{simon_2007}.  

Located near the North Galactic Pole at a distance of $224$~kpc from
the Earth \citep{zucker_2006}, the Canes Venatici~I dwarf spheroidal
galaxy (CVn~I dSph, $\alpha_0 =13^h28^m03.5^s, \delta_0 =
33^\circ33'21.0''$) was, at the moment of its discovery, one of the
most remote satellites to the MW \citep{zucker_2006}.  Being a
relatively luminous companion to the MW ($M_v = -7.9 \pm 0.5$ mag
\cite{zucker_2006}), its large distance allowed it to remain
undetected until 2006.  With a half-light radius of $564$~pc
\citep{zucker_2006}, it is one of the largest satellites of the MW
\citep{mcconnachie_2012,zucker_2006}.  It possesses an old ($\sim
12$~Gyr, \cite{okamoto_2012}), metal-poor ([Fe/H]$\sim -2$
(\cite{zucker_2006}, \cite{simon_2007}) population of stars that
represents 95\% of the luminous mass of the galaxy, and a younger
($\sim 1.4-2.0$~Gyr) more metal-rich ([Fe/H]$\sim -1.5$, ) population
of stars, revealed by a blue plume in the galaxy's colour-magnitude
diagramme \citep{martin_2008}, accounting for the remaining 5\% of the
stars.

As commonly observed in this type of galaxies \citep{mateo_1998},
CVn~I has a high velocity dispersion, with a value of $7.6$
km\,s$^{-1}$ \citep{simon_2007}.  If one assumes virial equilibrium,
spherical symmetry, and isotropic velocity dispersion, this value
indicates a virial mass of $2.7 \times10^7\, \mathrm{M_\odot}$  and,
therefore, a $M/L \sim 220$ \citep{simon_2007}, suggesting that the
internal dynamics of this galaxy are dominated by a DM halo.  The
line-of-sight velocity of CVn~I is given in \citet{simon_2007} and
amounts to $30.9 \pm 0.6$~km\,s$^{-1}$.  In a recent publication
\citet{fritz_2018} used the Gaia DR2 data to determine a possible
proper motion of the dwarf.  They calculate $\mu_{\alpha} = -0.159 \pm
0.094 \pm 0.035$~mas\,yr$^{-1}$ and $\mu_{\delta} = -0.067 \pm 0.054
\pm 0.035$~mas\,yr$^{-1}$. 

As noted on different occasions \citep{zucker_2006, okamoto_2012,
  martin_2008b}, the CVn~I dSph has an elongated shape, with a
ellipticity of $\epsilon = 0.38$.  In figure 8 of
\citet{okamoto_2012}, we see that it is a very flattened and 
elongated system.  Part of this elongation may be due to the onset of
tidal tails, that is the object would be heavily influenced by Galactic
tides. 

It has been shown that tidally disrupted objects have their
line-of-sight velocity dispersions boosted by unbound stars that
form the tidal tails  \citep{read_2006b, munoz_2008, klimentowski_2009,
  smith_2013c, blana_2015}.  This boost is strongest if the object is
observed near its apo-centre, where objects spend most of their
orbital time according to Kepler's second law.  This increased value,
under the assumption of virial equilibrium, leads to an overestimation
of the dynamical mass of the system and to an elevated M/L.

According to some astronomers, the presence of a disc of satellites
around the MW \citep{pawlowski_2012} and M31 \citep{ibata_2013}
suggests that each of these galaxies may have suffered a major
interaction at some point in their past \citep{sawa_2005,
  pawlowski_2011, fouquet_2012, hammer_2013}.  One possibility is that
CVn~I might have been formed as a small tidal dwarf galaxy (TDG;
\citealt{duc_2011}) in such an interaction.  Simulations show that
these second generation objects can form from tidal interactions
between galaxies \citep{wetz_2007, bournaud_2008,  ploeckinger_2018}
and survive the initial stages of their formation \citep{reicchi_2007,
  ploeckinger_2014} up to several Gyr \citep{bournaud_2003}.  Also,
observations indeed show that not only are these kinds of objects
formed in galaxy interactions \citep{kaviraj_2012,scott_2018}, but
they can survive several Gyr around their progenitors
\citep{duc_2014}.  If that is the case, then Canes Venatici~I might
not be a DM-dominated object, since the gravitational collapse of the
material that forms a tidal tail produces an object that is unable to
capture any DM particle \citep{bournaud_2010}.  

A DM dominated dSph must lose $\sim 90$\% of its dark matter halo
before the luminous part can be affected by the gravitational
influence of the MW \citep{smith_2013a}.  If Canes Venatici~I is
affected and elongated by tides, then even though it would have been 
DM dominated in the past, DM is not the cause for the high velocity
dispersion we see today.

Dwarf disc interactions \citep{mayer_2007,d'onghia_2009} work for
larger dwarfs, that is reducing the stellar content and reshaping it
while at the same time leaving the DM halo more or less intact, but
faint and ultra faint dSph may not have formed from dwarf discs as
they have very low mass haloes with circular velocities in the order
of or below the velocity dispersion expected (Hazeldine \& Fellhauer,
2018, private communication).

\begin{table}
\caption{Observable properties of CVn~I.  References: (1)
  \citet{zucker_2006}, (2) \citet{martin_2008b}, (3)
  \citet{simon_2007} } 
  \label{obs}
  \begin{tabular}{llrc}
      \hline
      Observable & & Value & \\
      \hline
      Right ascension & $\alpha_0$ & $13^h 28^m 03.5^s \pm 1.3$ & (1) \\
      Declination & $\delta_0$ & $33^\circ 33' 21'' \pm 10$ & (1) \\
      Distance & $D$ & $224 \pm 10$~kpc & (1) \\
      Ellipticity & $\epsilon$ &  $0.38 \pm 0.03$  & (1) \\
      Position Angle & PA & $73^\circ \pm 3^\circ$ & (1) \\
      Total Luminosity & $L_{V}$ & $2.3 \pm 0.3 \times 10^5$ L$_\odot$ &
      (2) \\ 
      Cent. surf. bright. & $\mu_0$ & $27.1 \pm
      0.2$~mag\;arcsec$^{-2}$ & (2) \\ 
      Half-light radius & $r_{\rm h}$ & $564 \pm 36$~pc & (2) \\ 
      Radial velocity & $v_{\rm rad}$ & $30.9 \pm 0.6$~km\;s$^{-1}$
      & (3) \\ 
      Velocity Dispersion & $\sigma_{\rm los}$ & $7.6 \pm
      0.4$~km\;s$^{-1}$ & (3) \\ 
      \hline 
    \end{tabular}
\end{table}

The aim of this project is to find a possible DM free progenitor, able
to reproduce the observed properties of CVn~I mentioned above and shown
in Tab.~\ref{obs}.  In the next section, we explain the setup of our
simulations followed by our results.  We end this paper with some
conclusions and briefly discussing our results. 

\section{Setup}
\label{sec:setup}

\subsection{Infall time}
\label{sec:infall}

The 'infall time' denotes the starting point of the simulations.  As
dSph have an old and metal-poor population of stars, meaning that they
stopped creating new ones long time ago, this starting point should
reflect the time when the majority of the stars were created.  
Spectroscopic data shows that CVn~I dSph is dominated by an old and
metal-poor stellar population \citep{zucker_2006}, and isochrone
fitting indicates that the age of these stars is at least $10$~Gyr old,
with a younger population of $\sim1.2$-$2.0$~Gyr that makes up to
$\sim 5$\% of the mass of the galaxy \citep{martin_2008}. 
We use this value as a reference to chose a generic 'infall time' of
$10$~Gyr, to account for the fact that we do not know the true value
for this parameter.   

We use the term 'infall time' for the start of our simulations, even
though it might be misleading.  If CVn~I was indeed originally a DM
dominated dwarf galaxy, the start of our simulations marks the point
in its evolution, where most of the DM halo ($> 90$\%) was already
stripped away in the past.  In the case that CVn~I was DM free from
its formation, we might be looking at a tidal dwarf galaxy which has
formed orbiting the MW instead of falling in. 

Previous studies \citep[e.g.][]{blana_2015} show that a change in
'infall time' (e.g.\ from $10$ to $5$~Gyr) does not alter our
conclusions, that is it is possible to reproduce the observables of 
a dSph with a DM-free object.  It just changes the properties of the
progenitor, which can be less massive or less concentrated to begin
with, as it needs less time to survive the tidal forces of the MW. 

\subsection{Orbit candidates}
\label{sec:orbit}
    
The trajectory of a particle in the potential of the MW is completely
defined if one knows the 3D position and the 3D velocity vectors. 
Using a point mass integrator (PMI) it is possible to integrate
backwards in time to obtain the path of the galaxy and its position
and velocity $10$~Gyr ago, with the resulting values as initial
conditions at the beginning of the simulation, using potentials to
model the MW following \citet{mizutani_2003}.

The galactic disc was modelled using the Miyamoto-Nagai
\citep{miyamoto_1975} profile, defined as:  
\begin{equation}
  \label{M-Nagai}
  \Phi^{\rm MN}_{\rm disc}(R,z) =
  \frac{-GM_{\mathrm{disc}}}{\left(
      R^2+\left[a+(z^2+b^2)^{1/2}\right]^2\right)^{1/2}} 
\end{equation} 
where the parameters are $M_{\mathrm{disc}} = 10^{11}$
$\mathrm{M}_\odot$, $a=6.5$~kpc and $b=0.26$~kpc.  To describe the
bulge of the MW, the Hernquist model \citep{hernquist_1990}  was used: 
\begin{equation}
  \label{Pl}
  \Phi_{\mathrm{bulge}}^{\mathrm{Hernquist}}=-\frac{GM_{\mathrm{bulge}}}{r + c}
\end{equation}
with $c=0.7$~kpc and $M_{\rm bulge}= 3.4 \times 10^{10}$~M$_{\odot}$.
At the distance where CVn~I dSph is currently located, the DM halo of
the MW is nearly spherical \citep{fellhauer_2006}, so a logarithmic
halo was chosen to model this structure: 
\begin{equation}
  \label{loghalo}
  \Phi^{\mathrm{log}}_{\mathrm{halo}} = v_0^2\ln(r^2+d^2)
\end{equation}
where the values of the parameters are $v_0 = 131.5$~km\,s$^{-1}$ and
$d = 12$~kpc. 
    
For CVn~I dSph we know its position in the sky, its distance to the
Sun and its velocity in the line-of-sight, that is the three
components of its position vector and only one component of the
velocity vector.  The two proper motions of the galaxy as seen from
Earth are still needed to compute a possible orbit: along the
declination axis ($\mu_{\delta}$) and along the right ascension axis
($\mu_{\alpha}$).  
  
\begin{figure}
  \centering
  \includegraphics[width=\columnwidth]{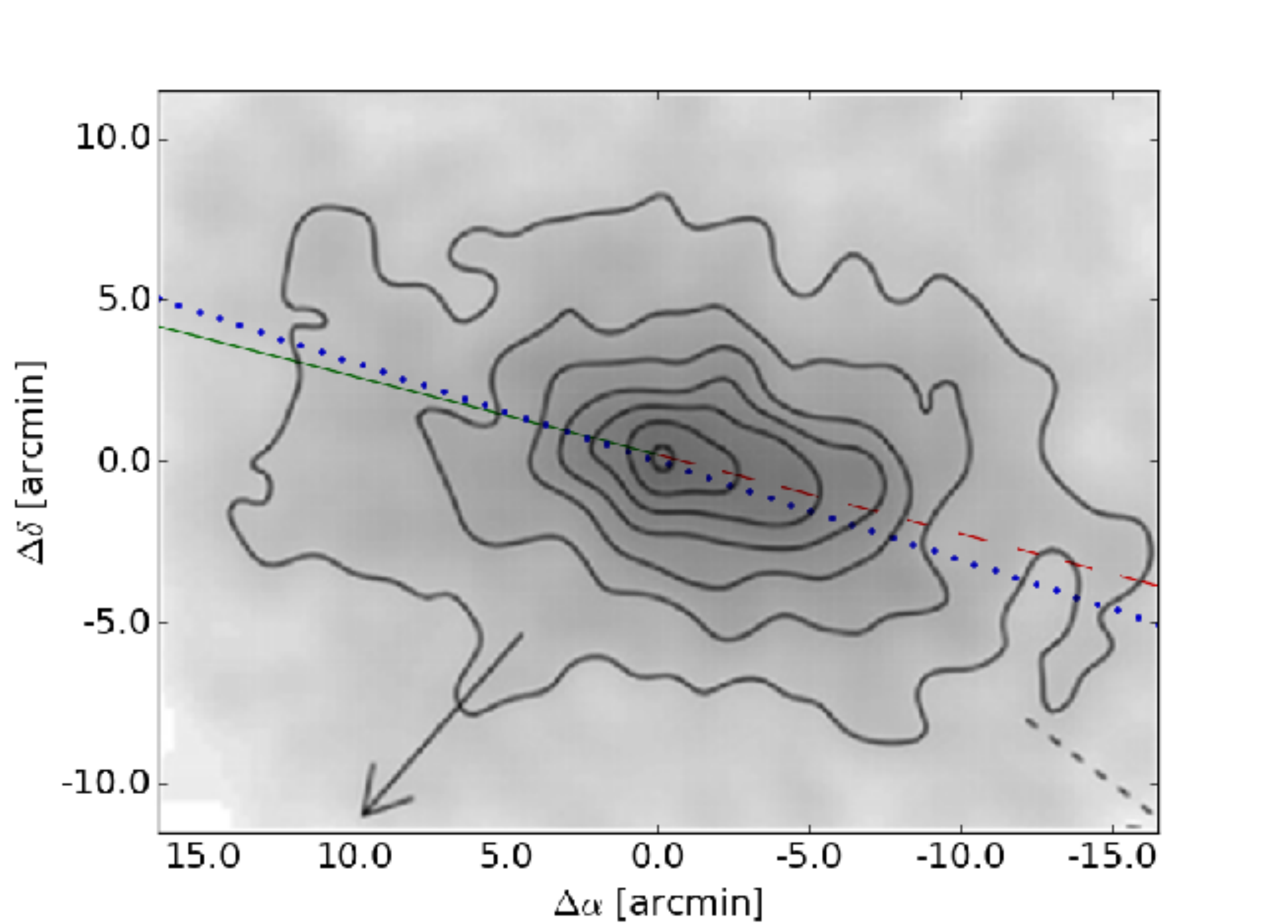}
  \caption{Projected orbit of CVn~I dSph on the sky. The red dashed
    line represents the past position of CVn~I dSph, while the green
    line corresponds to the future position of the galaxy. The blue
    dots are aligned to the major axis of the dwarf. The isodensity
    contour plot was adapted from \citet{okamoto_2012}, figure 8.} 
  \label{fig:proy_orb}
\end{figure}

Even though there is an infinite number of pairs $(\mu_{\alpha},
\mu_{\delta})$, using the fact that galaxies that are not DM
dominated (either because they lost their DM or were born without
it) and are undergoing tidal disruption will show tidal tails along
their orbit, leaves us with only those pairs of proper motion,
generating orbits parallel to the projected major axis of the
elongated dSph.  In Fig.~\ref{fig:proy_orb} we show the past
position of the test particle as the red line and the future
position as the green line.

That tidal tails are aligned with the orbit is true if we see the
object close to its apo-centric distance.  On an eccentric orbit,
objects lose stars while going through peri-centre or due to tidal
shocks passing the disc of the main galaxy.  The particles (stars)
leave the object through the Lagrangian points L1 and L2, that is
perpendicular to their orbits.  These newly lost particles will align
with the orbit on the way out to the apo-centre.  

Old tails stay aligned with the orbit, which may lead to strange
appearances, closely after peri-centre, of 4 tails, a phenomenon which
\citet{blana_2015} dubbed 'X-wing' shape.  A similar behaviour is
quoted in \citet{klimentowski_2009}, even though they are using a
dwarf disc galaxy embedded in a dark matter halo.  They see that the
new tidal tails emerge perpendicular to the orbit after the
peri-centre passage and align with the orbit during the later
evolution.  In contrast to objects like star clusters or in general
spherical, dispersion supported objects without dark matter, like in
this study, they discuss that in their case the alignment needs longer
time and their tails are still perpendicular at apo-centric distances.
The velocity dispersion of the galaxy is boosted when it reaches
apogalacticon  \citep{smith_2013c}, so a second condition that the
final orbit must satisfy is that the apogalactic distance must be
close to the current distance of CVn~I dSph. 

\begin{figure}
  \centering
  \includegraphics[width=\columnwidth]{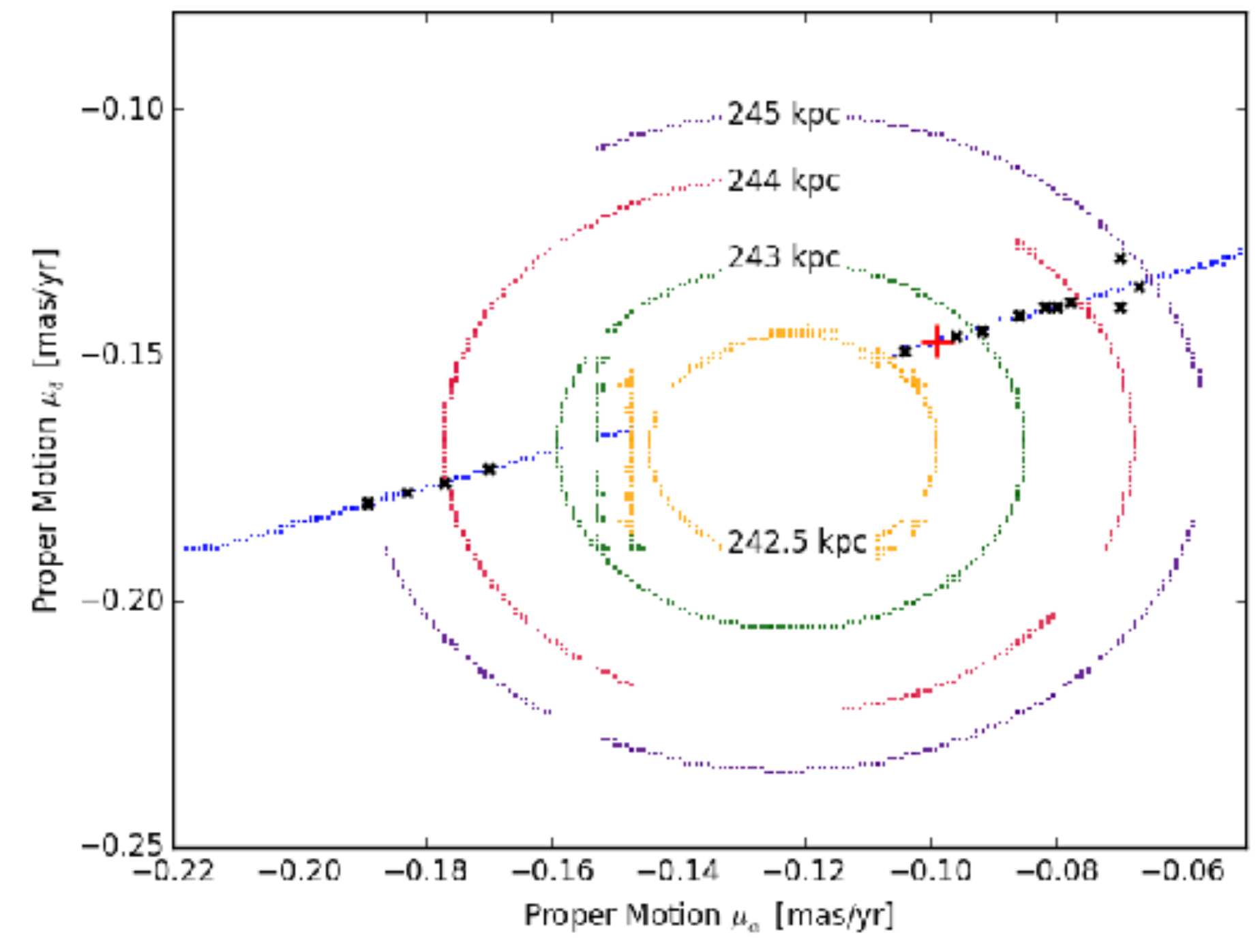}
  \caption{Proper motions of the orbit candidates. The blue dots
    represent the pairs of proper motions that match the position
    angle of CVn~I, while the purple, red, green, and yellow dotted
    circles mark the zone of parameter space where the orbits have
    apogalactic distances of $245$~kpc, $244$~kpc, $243$~kpc, and
    $242.5$~kpc, respectively. The black crosses are the candidates
    for the final orbit (Tab.~\ref{orbitas}), while the red plus sign
    is the selected orbit.} 
  \label{fig:cand}
\end{figure}

\begin{table}
  \centering
  \caption{List of candidates for the CVn~I dSph orbit. Each of these
    candidates were chosen taking in account that the orbit must be
    parallel to the mayor axis of the galaxy as seen from Earth and
    its apogalactic distance must be similar to the current
    galactocentric distance of CVn~I dSph, which is $D_\odot = 224$
    kpc \citep{simon_2007}} 
  \label{orbitas}
  \begin{tabular}{c c c c c}
    \hline
    No. & $\mu_{\alpha}$ & $\mu_{\delta}$ & Pericentre & Apocentre \\
    & [mas\;yr$^{-1}$] & [mas\;yr$^{-1}$] & [kpc] & [kpc] \\
    \hline
    1 & -0.104 & -0.149 & 16.687 & 242.632 \\
    2 & -0.099 & -0.147 & 20.013 & 242.790 \\
    3 & -0.170 & -0.173 & 20.029 & 243.596 \\
    4 & -0.096 & -0.146 & 22.312 & 242.916 \\
    5 & -0.177 & -0.176 & 24.989 & 244.049 \\
    6 & -0.092 & -0.145 & 25.020 & 243.073 \\
    7 & -0.086 & -0.142 & 29.906 & 243.398 \\
    8 & -0.183 & -0.178 & 29.976 & 244.551 \\
    9 & -0.082 & -0.140 & 33.220 & 243.657 \\
    10 & -0.189 & -0.180 & 34.968 & 245.099 \\
    11 & -0.080 & -0.140 & 35.016 & 243.784 \\
    12 & -0.078 & -0.139 & 36.575 & 243.910 \\
    13 & -0.070 & -0.140 & 42.739 & 244.398 \\
    14 & -0.067 & -0.136 & 46.355 & 244.808 \\
    15 & -0.070 & -0.130 & 45.675 & 244.855 \\
    \hline 
  \end{tabular}
\end{table}

The possible pairs of proper motions are selected from the
intersection of the regions, found using the method described in
Appendix \ref{sec:ap1}, that satisfy the criteria mentioned above.  
These  candidates are summarised in Tab.~\ref{orbitas} and shown in
Fig.~\ref{fig:cand}. 
The blue dotted line represents the pairs of proper motions that match
the position angle of CVn~I, while the purple, red, green, and yellow
dotted circles mark the zone of parameter space where the orbits have
apogalactic distances of $245$~kpc, $244$~kpc, $243$~kpc, and
$242.5$~kpc, respectively.  The black crosses are the candidates for
the final orbit (Tab.~\ref{orbitas}), while the red plus sign is the
selected orbit.  Orbits in the centre of the concentric circles have
the smallest apogalactic and perigalactic distance, with $242.2$~kpc
and $7.8$~kpc, respectively, but they do not match the position angle
of the galaxy.  

For each of these candidates we run one simulation for $10$~Gyr,
using a Plummer sphere with a scale length (Plummer radius $R_{\rm
  pl}$) of $630$~pc and a total mass of $M_{\rm pl} =
2.13\times10^7$~M$_{\odot}$, distributed into one million particles.
We use \texttt{Superbox} \citep{superbox}, a particle-mesh code,
where a sub-grid with a spatial resolution of $R_{\rm core} = R_{\rm
  Pl}/15$ was selected to ensure that these models are stable in
isolation.  In a particle-mesh code it is possible to use an
arbitrary number of particles as they represent phase-space elements
and not single stars.

\begin{figure}
  \centering
  \includegraphics[width=\columnwidth]{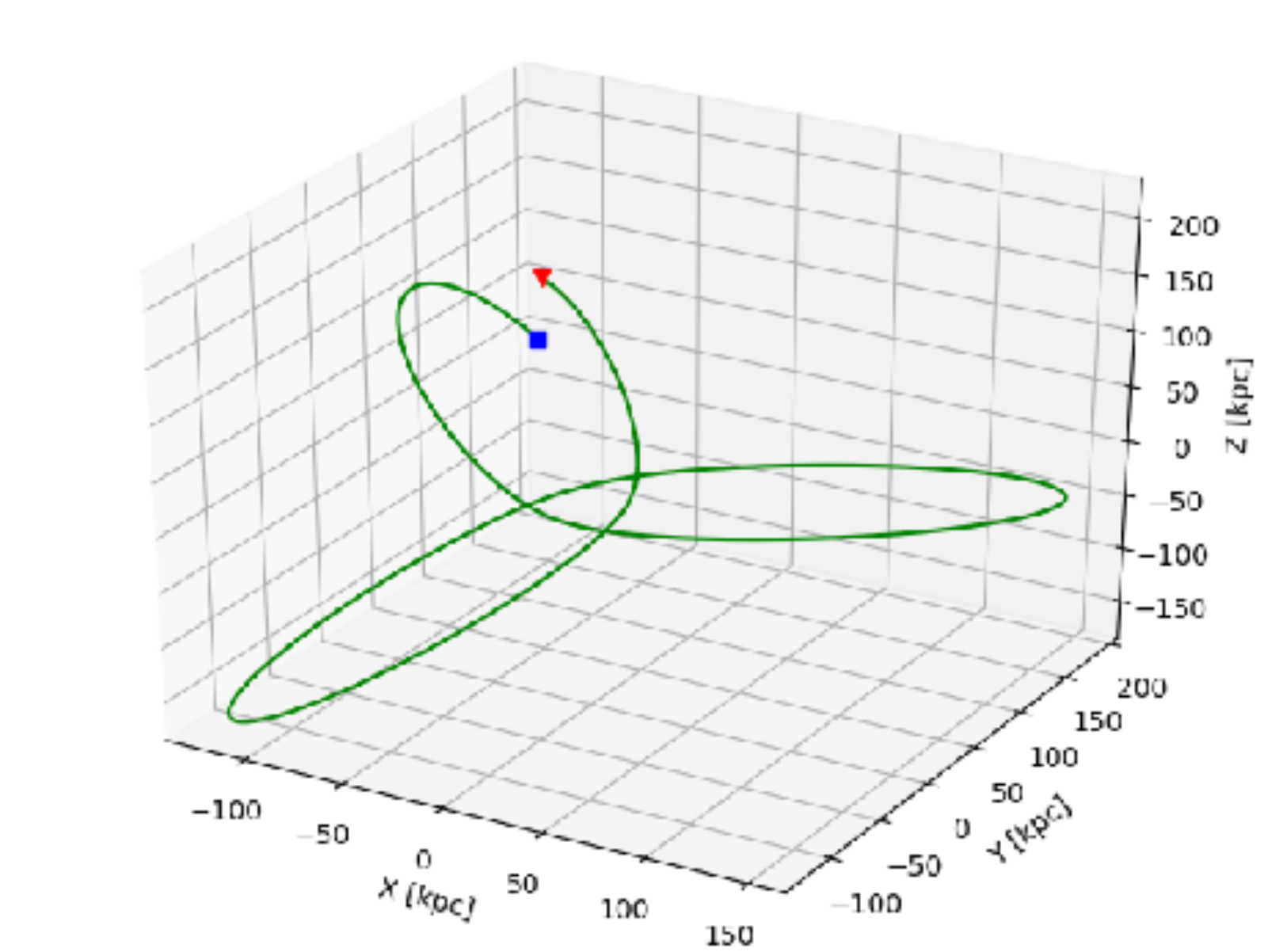}
  \caption{Selected orbit for CVn~I.  The green line shows the
    position of the point mass particle backwards in time, the
    blue square the position at $t=-10$ Gyr and the red triangle
    the current position of CVn~I.  Axes centred on the Galactic
    centre. } 
  \label{fig:orb3d}
\end{figure}

From the output of the test models, we choose the orbit whose
simulation ended with a remnant with values closest to the ones
observed in CVn~I dSph, so the orbit (2) from Tab.~\ref{orbitas} is
the one that we use.  This orbit has a perigalactic distance of
$20.013$~kpc and apogalactic distance of $242.79$~kpc, which gives us
an eccentricity of $\epsilon \sim 0.85$ and three pericentric passages
within the $10$~Gyr of simulation.  With our choice of proper motions
of $\mu_{\alpha} = -0.099$~mas\,yr$^{-1}$ and $\mu_{\delta} =
-0.147$~mas\,yr$^{-1}$ we are within the $1 \sigma$-errors of the
recently determined proper motions by \citet{fritz_2018} using the
Gaia DR2 data.  This is by coincidence as at the start of our study,
these values were not yet available.  The initial position and
velocity of the progenitor, according to the chosen orbit, are shown
in Tab.~\ref{tab:pos} and Fig.~\ref{fig:orb3d}. 

\begin{table}
  \caption{Parameters of the chosen orbit. These are the position and
    velocity of the particle determined by the PMI using the proper
    motions shown in the first two columns and then evolving back in
    time for $10$~Gyr.} 
  \label{tab:pos}
  \centering
  \begin{tabular}{l r r}
    \hline
    Orbit (2) & Today & Start \\
    t (Gyr) & 0 & -10 \\
    \hline
    x (kpc) & -2.698 & -7.430 \\
    y (kpc) & -38.113 & -34.172 \\
    z (kpc) & 220.474 & 164.306 \\
    vx (km\;s$^{-1}$) & -24.089 &  -40.174 \\
    vy (km\;s$^{-1}$) & -45.383 &  -70.832 \\
    vz (km\;s$^{-1}$) &  68.974 &  148.304 \\
    \hline
  \end{tabular}
\end{table}

\subsection{Initial conditions}
\label{sec:init}

We model our DM-free progenitor of CVn~I, using the chosen
orbit, again as a Plummer sphere \citep{plummer_1911} with similar
properties as the test model used to select the orbit: made of one
million, equal mass particles and using a sub-grid with spatial
resolution of $R_{\rm pl}/15$.  In contrast to the previous subsection
we now use the same orbit but we vary the parameters of the Plummer
sphere, that is the Plummer radius $R_{\rm pl}$ and the total mass of
the object $M_{\rm pl}$, to find the best matching final object.  A
total of 147 Plummer spheres were used, divided into 3 sets with 7
radii and 7 masses each, so that each set has 49 models.  A summary of
the properties of the used Plummer spheres can be found in
Tab.~\ref{tab:initial}.  

To reach the goal of finding the initial Plummer radius $R_{\rm pl}$ and
initial mass $M_{\mathrm{pl}}$ of a Plummer sphere that, after $10$~Gyr
orbiting the MW on the path chosen above, will generate a remnant that
reproduces the current observable parameters of CVn~I dSph, it is
necessary to understand how these observables are affected by
$R_{\mathrm{pl}}$ and $M_{\mathrm{pl}}$.  This will be explained in
more detail in the following section.

\begin{table}
  \centering
  \caption{Plummer radius and Plummer masses of each set of
    simulations.  Each model in a given set will use a different
    radius $R_{\rm pl}$ and mass $M_{\rm pl}$ taken from the
    respective list, giving a total of 49 models in each set.} 
  \label{tab:initial}
  \begin{tabular}{cc|cc|cc}
    \hline
    \multicolumn{2}{c|}{Set 1} & \multicolumn{2}{c|}{Set 2} &
    \multicolumn{2}{c}{Set 3} \\ 
    $R_{\rm pl}$ & $M_{\rm pl}$ & $R_{\rm pl}$ & $M_{\rm pl}$ &
    $R_{\rm pl}$ & $M_{\rm pl}$ \\ 
    $[\mathrm{pc}]$  & [$10^6 \mathrm{M}_\odot$] &$[\mathrm{pc}]$  &
    [$10^6 \mathrm{M}_\odot$] &$[\mathrm{pc}]$  & [$10^6
    \mathrm{M}_\odot$]\\ 
    \hline
    320 &  4.0  & 560 & 17.37 & 560 & 15.84 \\
    360 &  5.4  & 610 & 19.2  & 510 & 21.54 \\
    410 &  7.35 & 630 & 21.21 & 660 & 29.28 \\
    470 & 10.0  & 650 & 23.44 & 710 & 39.81 \\
    540 & 13.59 & 680 & 25.9  & 760 & 54.11 \\
    620 & 18.47 & 700 & 28.61 & 830 & 73.56 \\
    710 & 25.11 & 720 & 31.62 & 890 & 100.0 \\
    \hline
  \end{tabular}
\end{table}

\section{Results and final object}
\label{sec:res}

The remnants are analysed to find any trends that will help to
constrain the initial conditions of CVn~I dSph.   
We follow the method detailed in \citet{domingez_2016} and
\citet{blana_2015}.  In this method instead of trial and error to
get better and better matches of all observables simultaneously, we
search for functions of initial parameter sets which fit a certain
observable independently of the others.  The observational parameters
of the dwarf that we are trying to match can be found in
Tab.~\ref{obs}. 

\begin{figure}
  \centering
  \includegraphics[width=\columnwidth]{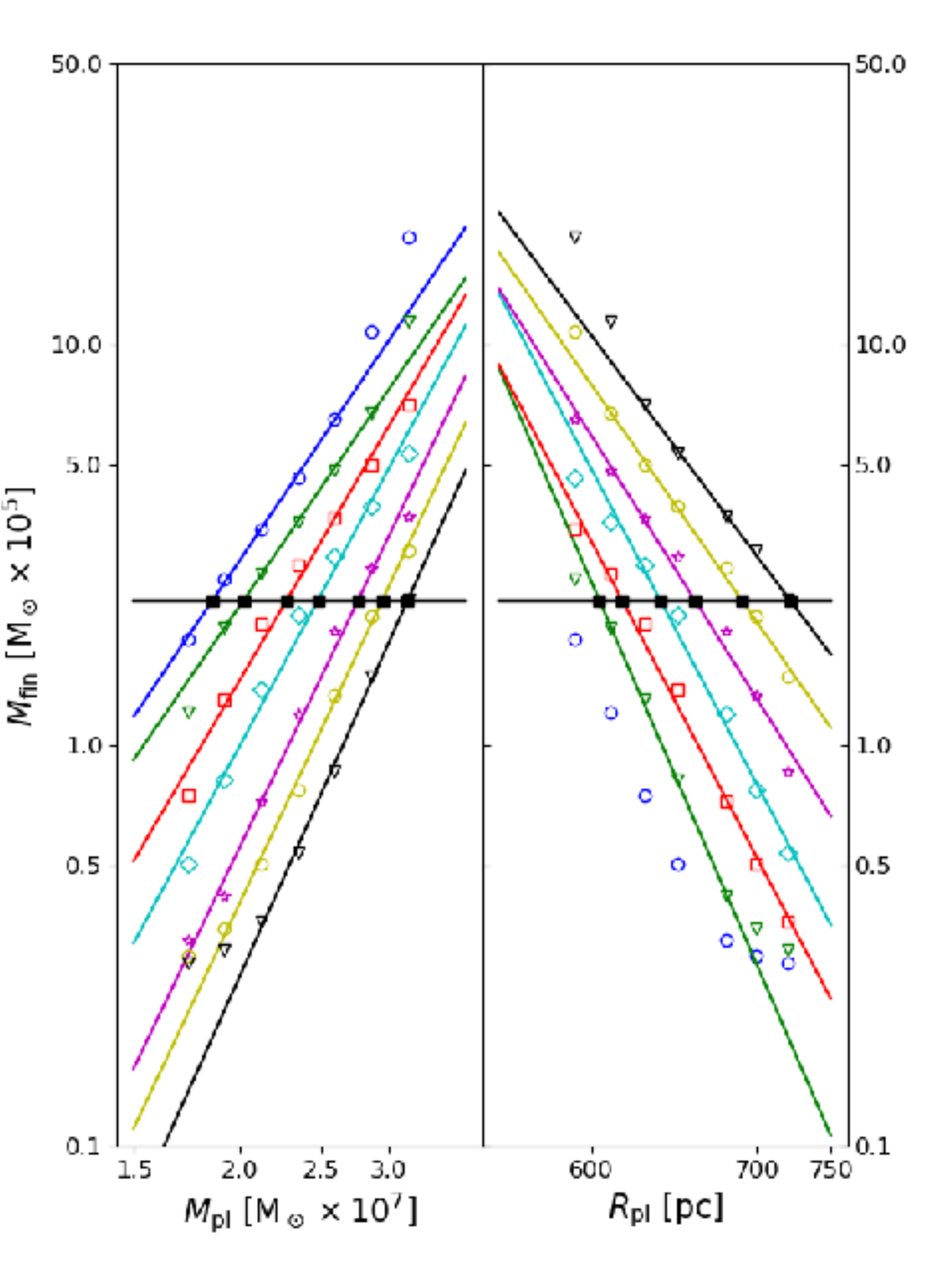}
  \caption{\small Final mass as a function of the initial parameters of
    the model. Left: Final mass as a function of the initial mass.
    Shown here are the data points and fitting lines for Plummer radii
    of 560 (blue dots), 610 (green triangles), 660 (red squares), 710
    (cyan diamonds), 760 (magenta stars), 830 (yellow circles), and
    890~pc (black triangles).  Right: Final mass as a function of the
    Plummer radius.  Shown here are the data point and fitting lines for
    Plummer masses of 17.37$\times 10^6$ (blue circles), 19.2$\times
    10^6$ (green triangles), 21.21$\times 10^6$ (red squares),
    23.44$\times 10^6$ (cyan diamonds), 25.9$\times 10^6$ (magenta
    stars), 28.61$\times 10^6$ (yellow circles), and 31.62$\times 10^6$
    M$_\odot$ (black triangles).  Horizontal solid line denotes the
    adopted value of the final mass that needs to be matched.  The
    black squares are the matching values calculated by fitting power
    laws to the data points.} 
  \label{fig:Mf}
\end{figure}

\begin{figure}
  \centering
  \includegraphics[width=\columnwidth]{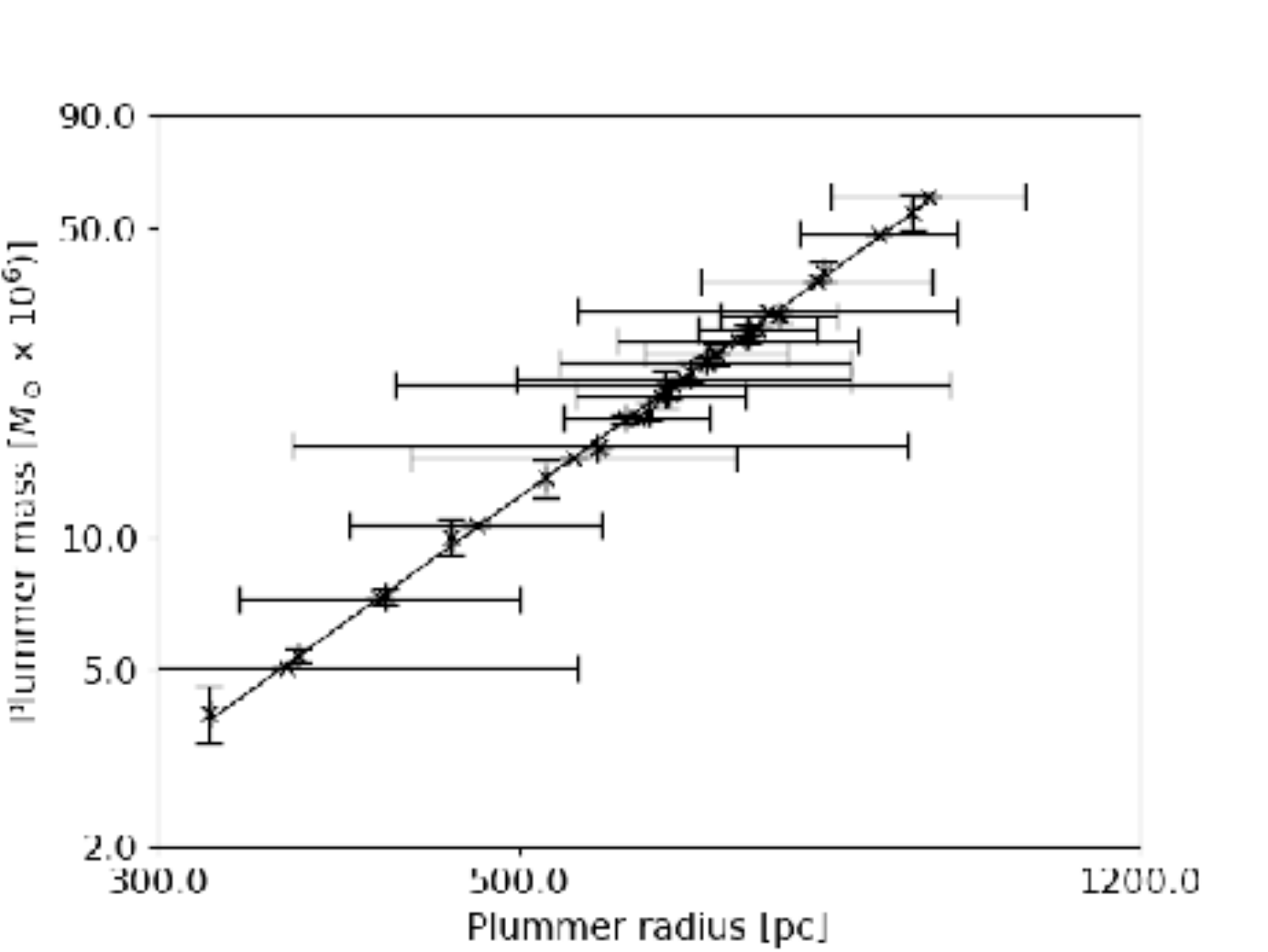}
  \caption{Pairs of initial parameters which lead to final models with
    the correct mass of Canes Venatici~I dSph.} 
  \label{fig:Mf-fit}
\end{figure}

When we are trying to fit a single observational parameter, we
only take its observational value into account and not the
observational errors that are given in Tab.~\ref{obs}.  As will be
shown in the following subsections, our mathematical errors due
to fitting procedures in logarithmic space are much larger that the
tight relations they accompany.  We believe that the extra
observational errors would only enlarge the error-bars by a small
factor. 

\subsection{Mass}
\label{sec:Mf}

The mass estimates for CVn~I dSph available in the literature were
calculated under the assumption that this object is spherical, is
in dynamical equilibrium, and has an isotropic velocity dispersion. 
From this, one gets a mass in the order of $2.7 \times
10^7\,\mathrm{M}_\odot$ and an M/L of $221$ \citep{simon_2007}, and,
therefore, a DM-dominated object.  Since we are trying to model this
galaxy as a DM-free object, we cannot use that value as the mass that
needs to be matched.   Instead, a generic mass-to-light ratio equal to
one is used to transform the observed luminosity of CVn~I dSph to an
estimate of its luminous mass. 

\citet{martin_2008b} reports that the total luminosity of CVn~I dSph
is $2.3\times 10^5\, \mathrm{L_\odot}$.  Using the M/L mentioned
above, the luminous mass of this galaxy (and the mass that our models
have to reach) would be $2.3 \times 10^5\, \mathrm{M_\odot}$.  
The mass calculated above uses all the light that comes from the
galaxy, that is not only the stars that are still gravitationally bound,
but also any tidal debris that might be in the vicinity. 
Following this principle, to measure the mass of the object at the
end of the simulation, we add the masses of all the particles inside a
box that covers $\pm 20'$ in DEC and $\pm 20'$ in RA around the centre
of density of the remnant, similar to the area observed by
\citet{martin_2008}. 

In Fig.~\ref{fig:Mf} we plot the measured final masses of our objects
as function of their initial $M_{\rm pl}$ (left panel) and $R_{\rm
  pl}$ (right panel) respectively.  In the left panel, all points with
the same symbol have the same value for $R_{\rm pl}$, showing how the
final mass depends on the initial mass for each choice of $R_{\rm
  pl}$.  As expected, a body with higher initial mass will produce a
more massive remnant, as the additional material will help to resist
the tidal influence of the MW.  In the right panel, all points with
the same symbol have the same value for $M_{\rm pl}$, showing how the
final mass depends on the Plummer radius.  Models with larger radii
lose more mass than the more compact ones since these are more loosely
bounded and more easily disrupted.   

For models with the same radius $R_{\rm pl}$, the points form a
straight line in this double logarithmic plot.  We fitted a power law
of the form $\log_{10} M_{\rm fin} = A \cdot \log_{10} M_{\rm pl} + B$
and find the value of $M_{\rm pl}$ for which $M_{\rm fin} = 2.3 \times
10^5\, \mathrm{M_\odot}$.  From this we obtain the pair $(R_{\rm
  pl},M_{\rm pl})$ that will generate a model that, at the end of the
simulation, will have a final mass with the desired value. 

\begin{figure}
  \centering
  \includegraphics[width=\columnwidth]{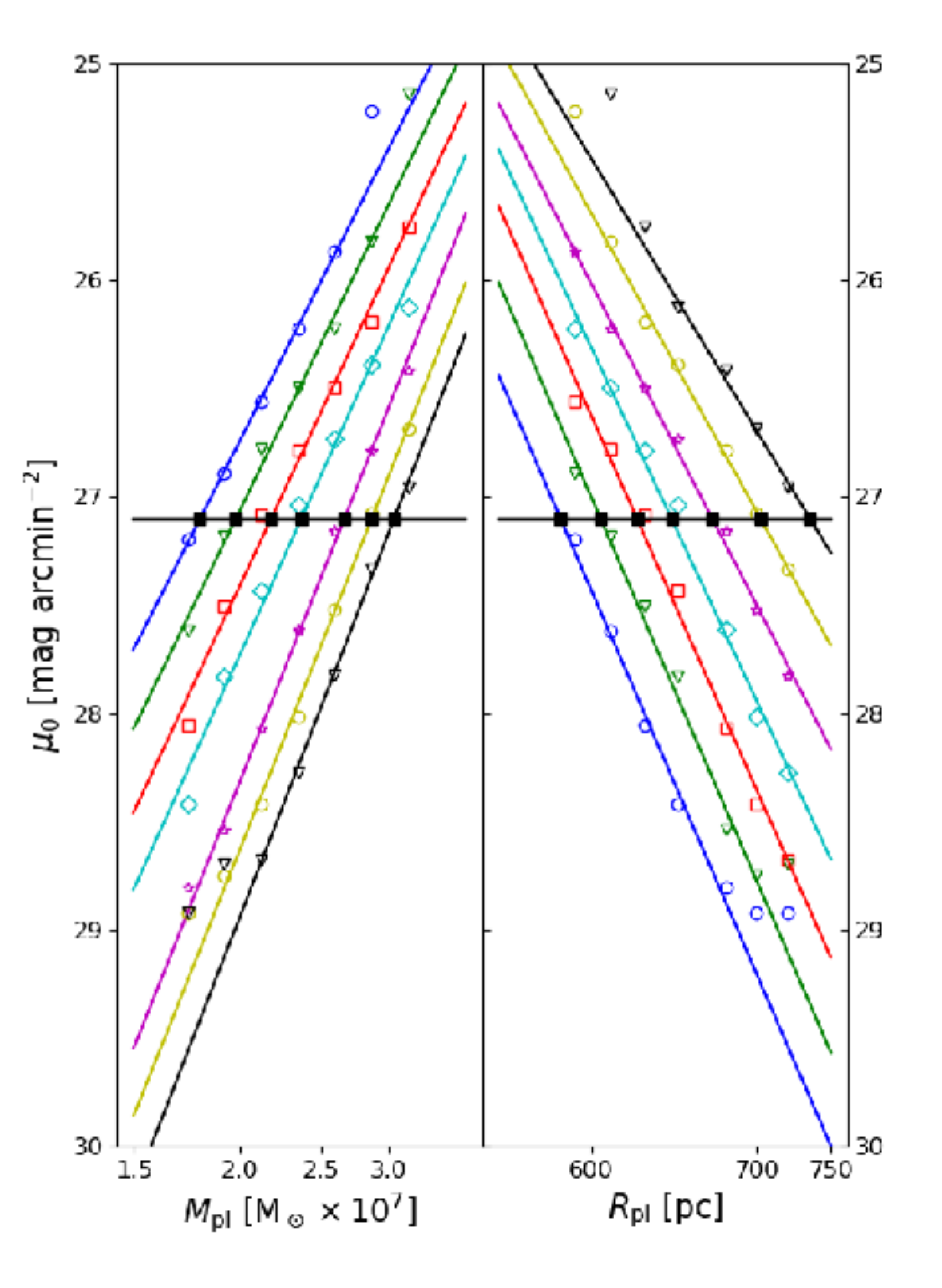}
  \caption{\small Same symbols as Fig. \ref{fig:Mf}, but for the
    central surface brightness. The horizontal black line denotes the
    observational value of 27.1.} 
  \label{fig:surf}
\end{figure}

\begin{figure}
  \centering
  \includegraphics[width=\columnwidth]{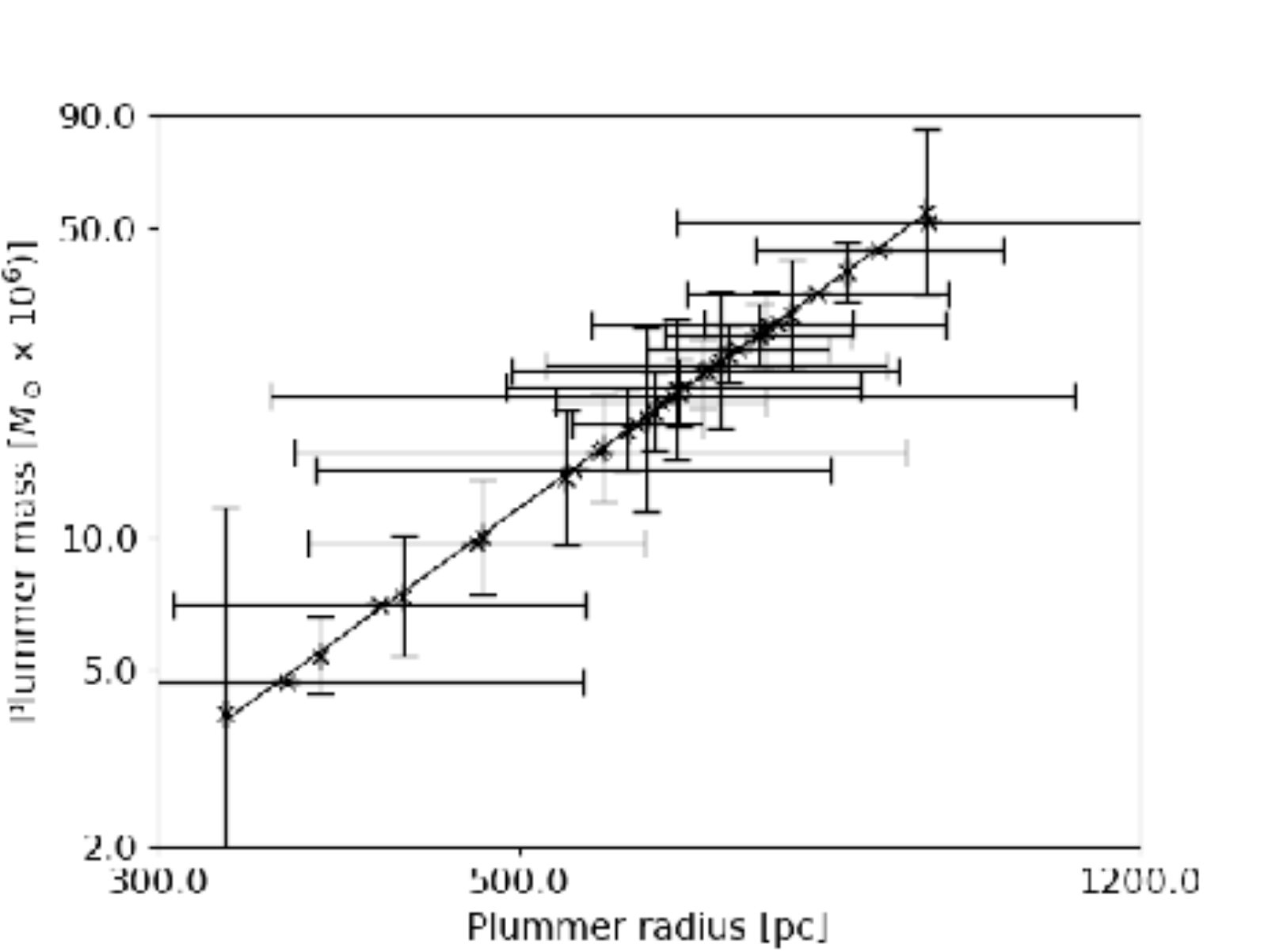}
  \caption{Pairs of initial parameters which lead to final models with
    the correct central surface brightness of Canes Venatici~I dSph.} 
  \label{fig:mag-fit}
\end{figure}

This is done for all sets with equal Plummer radius and then for all
the sets with equal Plummer mass (Fig.~\ref{fig:Mf}, right panel). 
These $(R_{\rm pl}$, $M_{\rm pl})$ pairs are, then, plotted in a
graph of Plummer mass vs Plummer radius with logarithmic axes
(Fig.~\ref{fig:Mf-fit}), where the points also form a straight line
that can be fitted with a power law. 
These points have error bars in only one direction, as the value in
one of the directions is given by the initial conditions of the
simulation, while the other stems from fitting a power law to a set of
points. 
Any points in this power law will produce a remnant with a final mass
close to $2.3 \times 10^5$.  The power law for this observable is: 
\begin{equation}
  \label{eq:Mf}
  M_{\rm pl}[\mathrm{M}_\odot] =  0.83 \pm 0.7 \times R^{2.65 \pm
    0.01}_{\rm pl}\, [\mathrm{pc}]. 
\end{equation}

The large error bars in this figure are due to the fitting
procedure to obtain the corresponding initial values.  A fitting
line from Fig.~\ref{fig:Mf} has an error in the zero-point and
another in the slope.  Both errors can affect the fitted value of
the searched for second initial parameter.  Even though we seemingly
have done a better job with our simulations, as the almost perfectly
aligned data points suggest, we give the correct mathematical
uncertainties.

Once again we stress that in the fitting routine we only consider
the observational value given in Tab.~\ref{obs} and pay no
consideration to the associated observational error.  Doing so would
only enlarge our already large mathematical errors.  The same
reasoning will be used in fitting the other observables as well.

\subsection{Surface brightness}
\label{sec:mag}

The same region used to measure the final mass of the object (see
Sec.~\ref{sec:Mf}) was used now to create a map of the surface mass
density with a resolution of $0.02^\circ$~px$^{-1}$.  The same generic
$M/L = 1$ was used to convert the surface densities (measured in
$[M_\odot \mathrm{pc}^{-2}]$) into surface brightness, measured in
magnitudes per square arcsecond.  The value of the brightest pixel was 
taken as the central surface brightness of the model.  The central
surface brightness of the models at the end of the simulation can be
found in Fig.~\ref{fig:surf}.  The value of the central surface
brightness that the final model has to reach is
$27.1$~mag\;arcsec$^{-2}$ (Tab.~\ref{obs}).   The models follow
similar trends that the ones described for the mass, were a more
massive object produces a more massive and, therefore, a brighter
remnant.  And a more extended object will lose more particles and end
up as a fainter object.  
        
To find the region of the space of initial parameters that yield a
model that matches the observed value, the procedure that was
described in Sec.~\ref{sec:Mf} is used.  The matching points in the
initial parameter space are shown in Fig.~\ref{fig:mag-fit}, where
they form a  tight power law that, after fitting an expression of the
form ${M_{\rm pl} = A \cdot \log_{10} R_{\rm pl} + B}$, leads to:  
\begin{equation}
  \label{eq:mag}
  M_{\rm pl}[\mathrm{M_\odot}] =  0.666_{-0.052}^{+0.056}\times
  R^{2.68\pm 0.01}_{\rm pl}\, [\mathrm{pc}] 
\end{equation}

\subsection{Effective radius}
\label{sec:reff}

To obtain the effective radius (defined as the radius of the isophote
containing half of the total luminosity) of the remnant, we used the
surface brightness as a function of the projected distance to the
centre of the object, and then fit a S\'ersic profile to get the
parameters.  To construct the light profiles, we omitted the central
$0.1^\circ$, which might still contain a bound object.  We fitted out to
$0.25^\circ$, the visible part of CVn~I dSph as seen by
\citet{okamoto_2012} and covering almost one kpc from the centre of 
the object, that is twice the effective radius measured by
\citet{martin_2008b}, using concentric circles to measure the light of
the remnant.  

\begin{figure}
  \centering
  \includegraphics[width=\columnwidth]{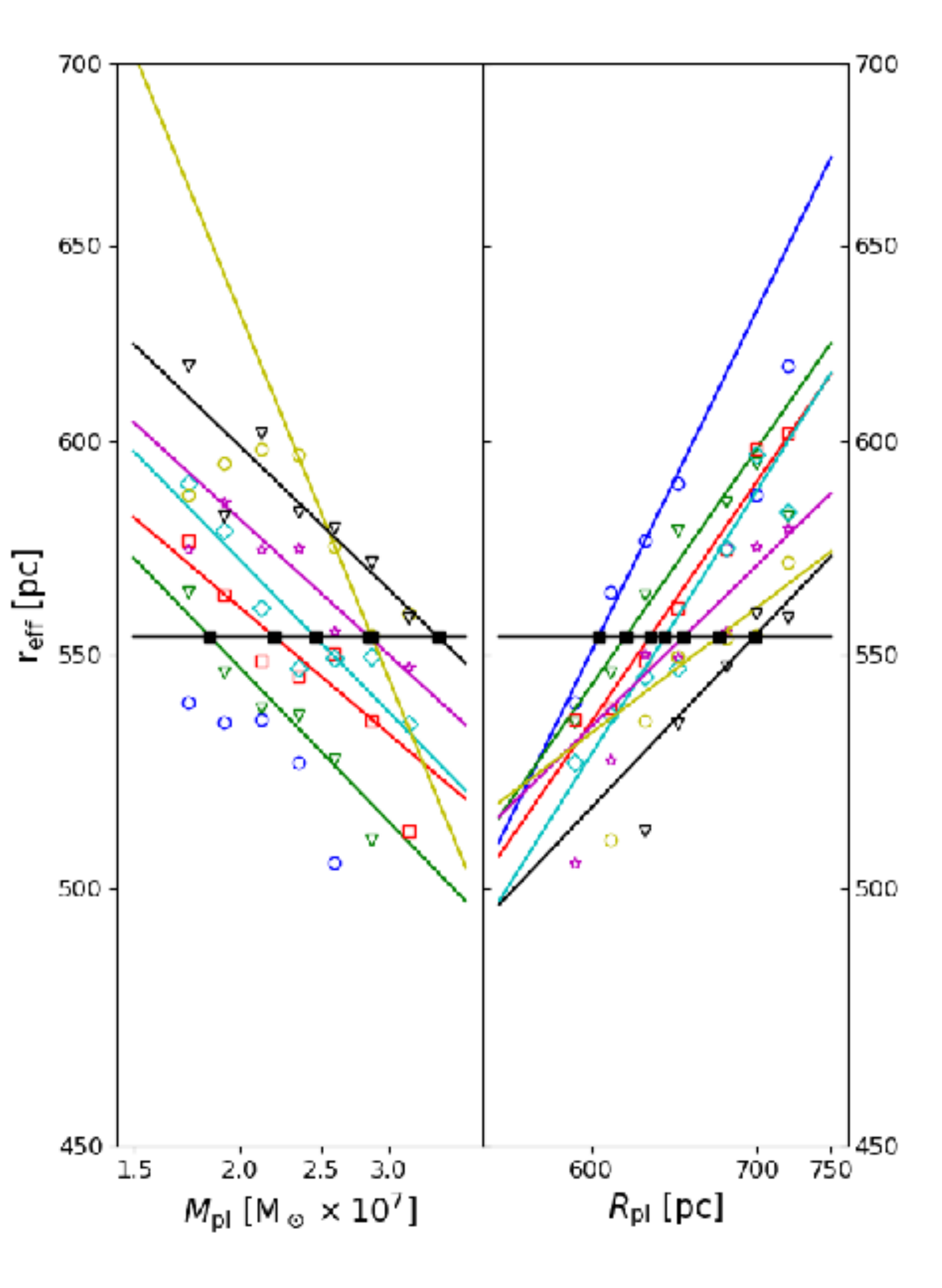}
  \caption{\small Same symbols as Fig. \ref{fig:Mf}, but for the
    effective radius. The horizontal black line denotes the
    observational value of 564 pc.} 
  \label{fig:reff}
\end{figure}
 
\begin{figure}
  \centering
  \includegraphics[width=\columnwidth]{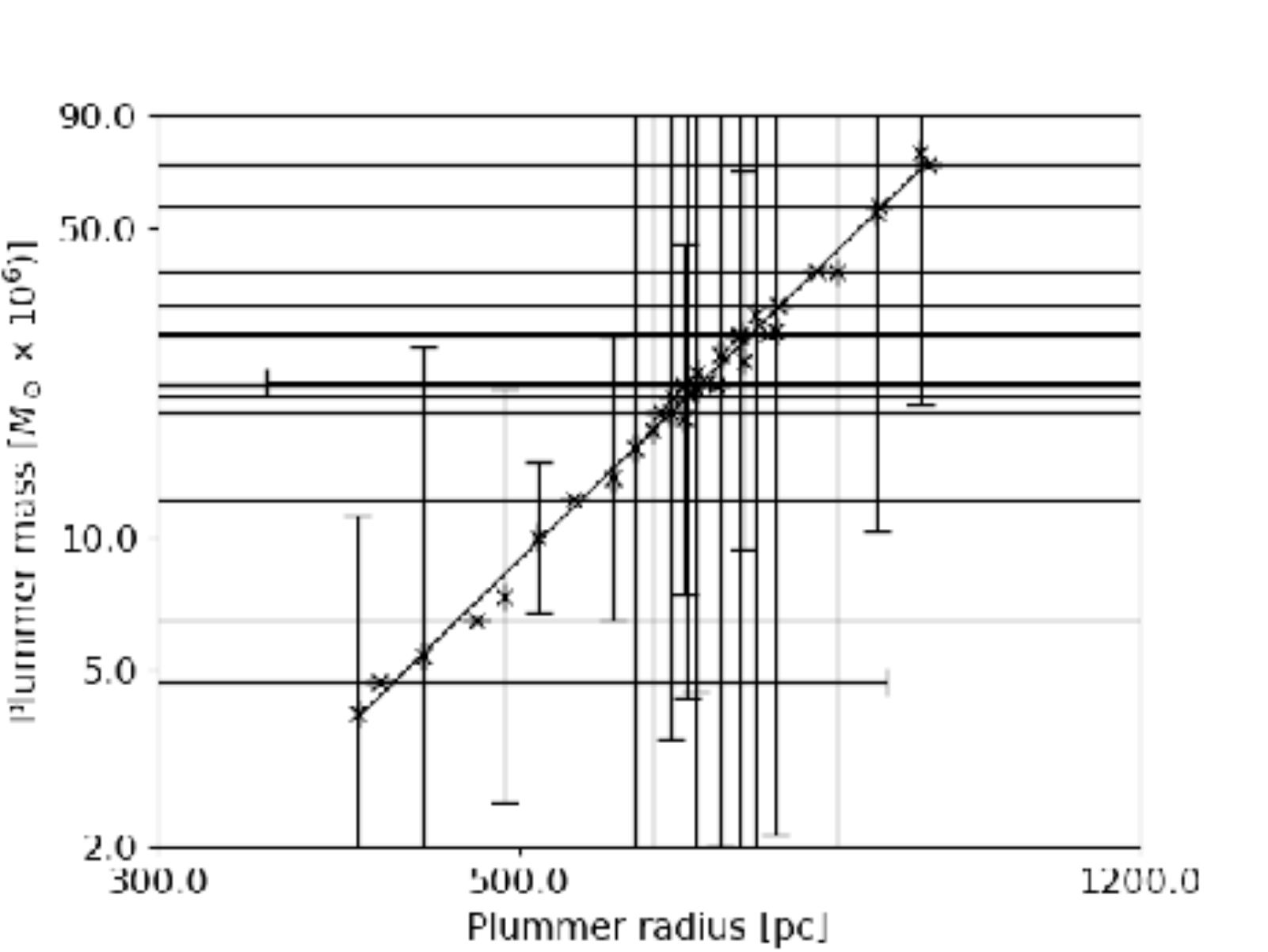}
  \caption{Pairs of initial parameters which lead to final models with
    the correct effective radius of Canes Venatici~I dSph. Despite the
    large error bars, the points form a tight power law, shown in
    Equation~\ref{eq:reff}.} 
  \label{fig:reff-fit}
\end{figure}

The effective radius of the models at the end of the simulation are
obtained via a S\'ersic fit and then plotted in a double logarithmic
graph (Fig.~\ref{fig:reff}).  From this plot we can see that for
compact objects (models with low $R_{\rm pl}$ or high $M_{\rm pl}$) we
obtain lower effective radii than for more extended and easily
disrupted Plummer spheres.  

\begin{figure}
  \centering
  \includegraphics[width=\columnwidth]{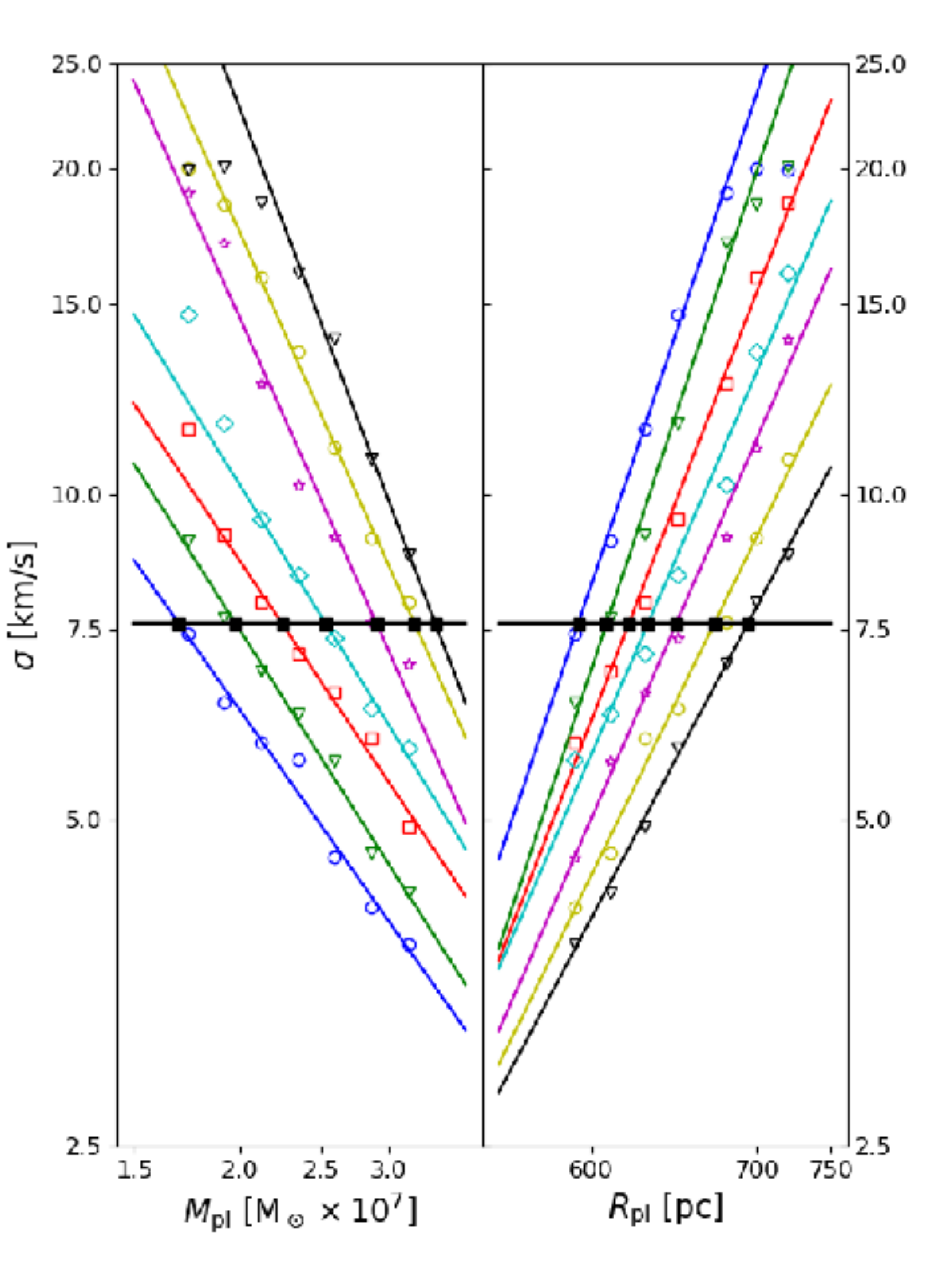}
  \caption{\small Same symbols as Fig.~\ref{fig:Mf}, but for the
    velocity dispersion.  The horizontal black line denotes the
    observational value of $7.6$~km\;s$^{-1}$.} 
  \label{fig:disp}
\end{figure}

\begin{figure}
  \centering
  \includegraphics[width=\columnwidth]{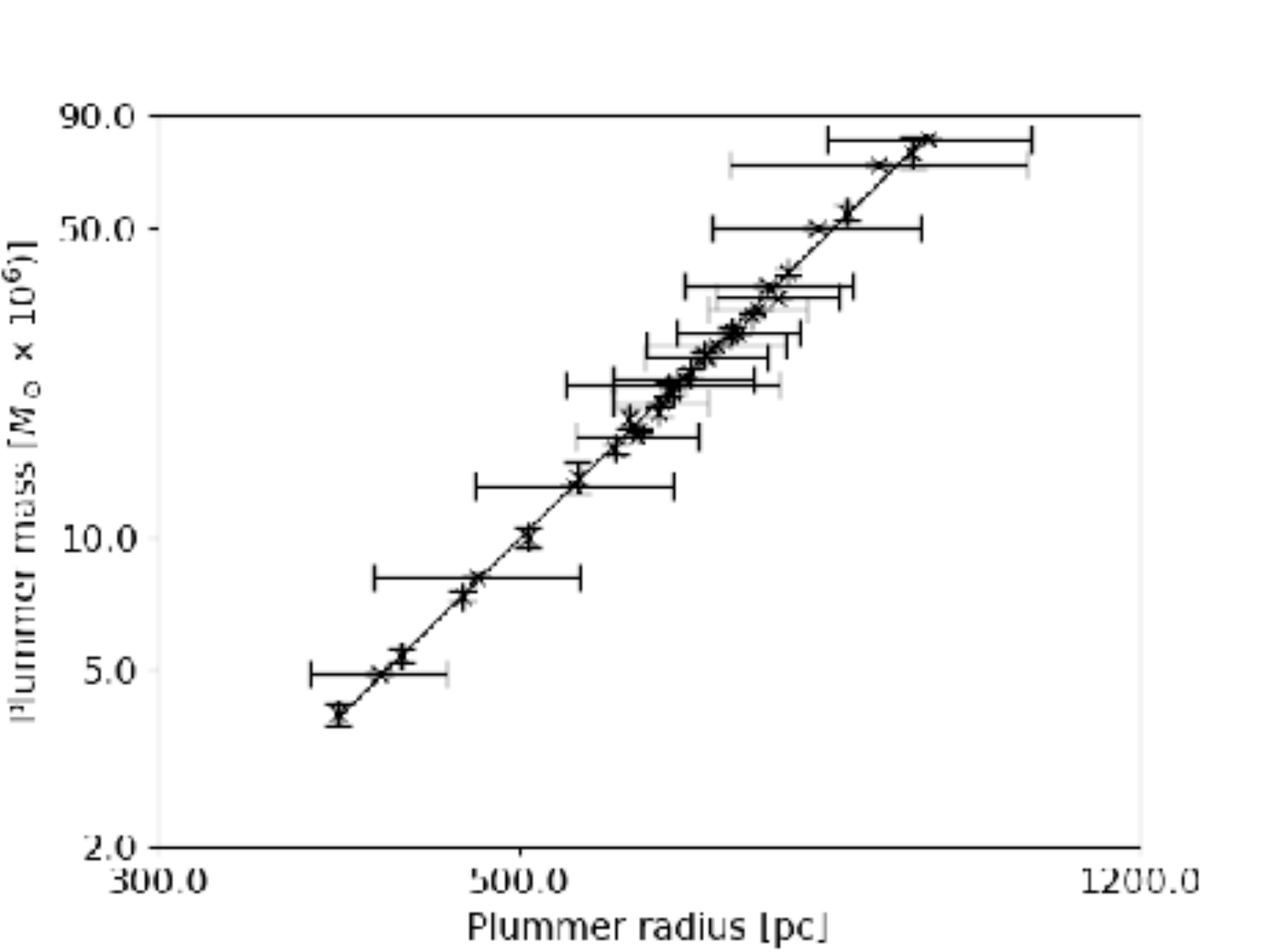}
  \caption{Pairs of initial parameters which lead to final models with
    the correct line-of-sight velocity dispersion of Canes Venatici~I
    dSph.} 
  \label{fig:sigma-fit}
\end{figure}

The value for the effective radius that we need to match with these
models is $564$~pc (see Tab.~\ref{obs}), represented with a black
horizontal line in Fig.~\ref{fig:reff}.  Following the same steps from
the previous sections we obtain the pairs of initial values that
reproduce the observable (Fig.~\ref{fig:reff-fit}) and fit the power
law to get:  
\begin{equation}
  \label{eq:reff}
  M_{\rm pl}[\mathrm{M}_\odot] =  0.0017_{-0.0005}^{+0.0007} \times
  R^{3.596\pm0.054}_{\rm pl}\, [\mathrm{pc}] 
\end{equation}

\subsection{Velocity Dispersion}
\label{sec:sigma}

To measure the velocity dispersion of the remnant, we use all the
particles inside a region covering $\pm 0.15^\circ$ in RA and $\pm
0.05^\circ$ in DEC from the centre of the object, equivalent to the
region used by \citet{simon_2007}, so we use not only the particles
from the bound centre, but also the ones that form the tidal tails of
the object.  They report a value of $7.6$~km\;s$^{-1}$ as the velocity
dispersion of CVn~I dSph, and we use this as the value that has to be
matched.  

Following the analysis made by \citet{blana_2015} we fit the solutions
where the object is almost destroyed, that is objects with low initial
masses and high Plummer radii, trying to match the velocity dispersion
mentioned above.  On this part of the parameter space, the central
part of the remnant, still bounded, is surrounded by a large number of
unbounded particles (the ones that form the tidal tails of the object
and can be found at the background and foreground of the centre of the
remnant).  

It is the influence of these particles plus the fact that for the
chosen orbit, at the end of the $10$~Gyr of simulation, the remnant is
close to the apocentre what boosts the velocity dispersion of the
object \citep{smith_2013c} to the observed levels, without using a
DM-dominated progenitor.  \citet{smith_2013c} have shown that for
the required boosting of the velocity dispersion the original object
has to be almost or even completely destroyed.  At this stage of the
evolution neither a $3-\sigma$ clipping nor the more elaborate IRT
method, quoted in \citet{klimentowski_2009}, are able to distinguish
between bound and unbound stars and return similar results as the
simple calculation used here \citep{smith_2013c}.

From Fig.~\ref{fig:disp}, where the black horizontal line represents
the observed value for the line-of-sight velocity dispersion, we see
that increasing the mass of the model does not increase the value of
this observable (left panel shows decreasing veloicty dispersion
with increasing mass; right panel has the fitting lines for the most
massive choices lower than for less massive objects, see the colour
coding and compare with previous figures), as would happen if the
object is in virial equilibrium, since a more massive object prevents
a larger loss of particles under the effect of the MW tides and,
therefore, there are less stars in the tidal tails that would increase
the value of $\sigma$.  A model with larger $R_{\rm pl}$ will lose more
mass and has more particles in the tidal tails, boosting the observed
velocity dispersion (left panel shows fitting lines for larger
initial objects higher up than for more concentrated one, compare
colour coding with previous figures; right panel shows increasing
velocity dispersion with increasing initial Plummer radius).
Again, after fitting the power laws from Fig.~\ref{fig:disp} and
obtaining the pairs of initial parameters that result in the correct
velocity dispersion, we plot them (Fig.~\ref{fig:sigma-fit}) and fit
the corresponding power law for this observable: 
\begin{equation}
  \label{eq:sigma}
  M_{\rm pl}[\mathrm{M}_\odot] = 0.0014_{-0.0002}^{+0.0003} \times
  R^{3.645\pm0.033}_{\rm pl}\, [\mathrm{pc}] 
\end{equation}

\subsection{Final model}
\label{sec:final}

\begin{figure}
  \includegraphics[width=\columnwidth]{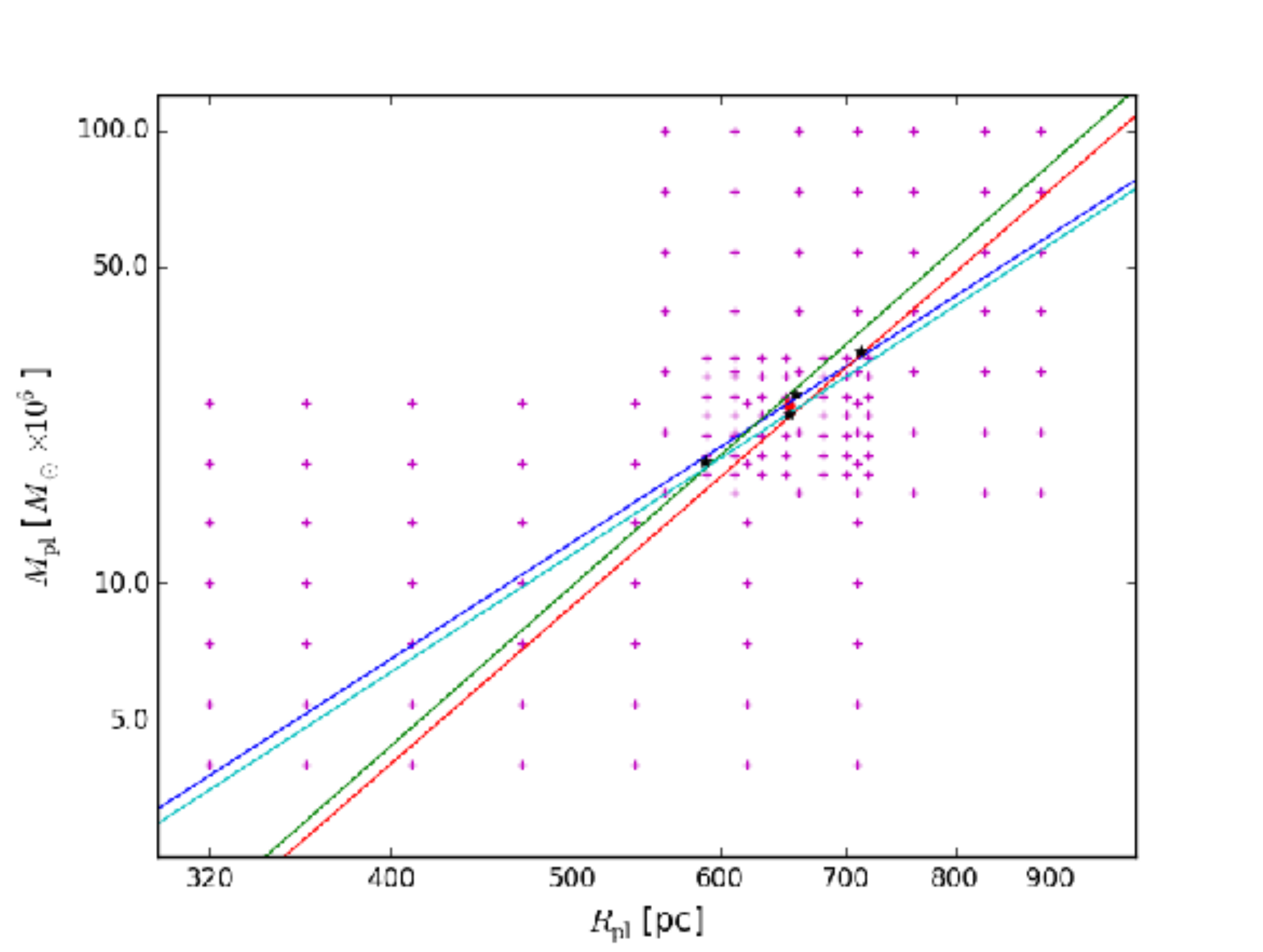}
  \caption{Zone where the best fitting model for CVn~I in the chosen
    orbit is located. Each line represents the initial parameters
    needed to match the four observables used on this project: Final
    mass (blue), central surface brightness (cyan), effective radius
    (red), and velocity dispersion (green). The magenta crosses are the
    parameters of the Plummer spheres used to calculate the power laws
    of Secs.~\ref{sec:Mf}, \ref{sec:mag}, \ref{sec:reff}, and
    \ref{sec:sigma}. The red circle is the final model and the black
    stars are the rejected models.} 
  \label{fig:final}
\end{figure}

\begin{figure}
  \centering
  \includegraphics[width=\columnwidth]{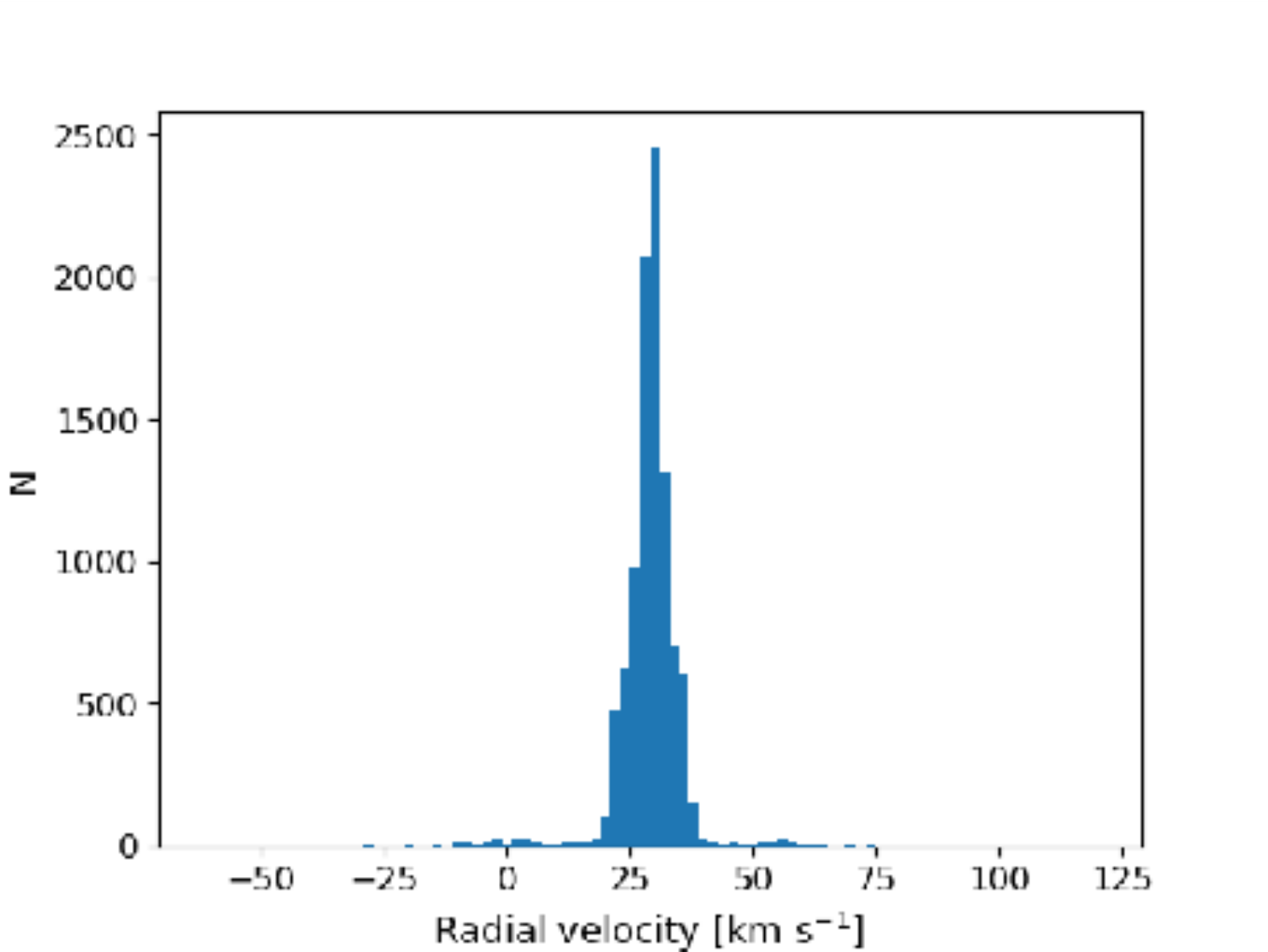}
  \caption{Velocity histogram of the final object.  We calculate the
    velocity dispersion of the remnant by selecting particles in the
    region covering $\pm 0.15^\circ$ in RA and $\pm 0.05^\circ$ in
    DEC.  Compare with figure~8 of \citet{simon_2007}} 
  \label{fig:hist_disp}
\end{figure}

The Eqs.~\ref{eq:Mf}, \ref{eq:mag}, \ref{eq:reff}, and
\ref{eq:sigma} represent the initial conditions that a Plummer sphere
must have, to match each of the observables, separately. 
Therefore, the intersection of these lines will indicate the pair
($R_{\rm pl}$,$M_{\rm pl}$) that matches all the observed properties
of CVn~I dSph.  In a ideal world, all the lines would intersect at the
same point of the parameter space.  In the real world, this does not
happen.  Nonetheless, the lines form a small region where the final
model, i.e\ the model that, after orbiting for $10$~Gyr in the orbit
determined in Sec.~\ref{sec:setup}, will produce a remnant whose
observational parameters are within the errors of the observational
data, is located.  In our case, the final model should be located in
the region with $600\, \mathrm{pc} \le R_{\rm pl} \le 700
\,\mathrm{pc}$ and $20.5\times10^6\,\mathrm{M_\odot} \le M_{\rm 
  pl} \le 30\times 10^6\,\mathrm{M_\odot}$. 

The four lines are shown in Fig.~\ref{fig:final}, together with the
initial parameters of the models used in this work.  In the area where
these four lines come close to each other, we use a trial and error
approach to find the final model.  A summary of the models used and
the final parameters at the end of the simulation can be found in
Tab.~\ref{tab:test}.  

\begin{table*}
  \centering
  \caption{\small Initial and final parameters of the candidates for
    the final model. Each column represents: Plummer Radius, Plummer
    Mass, Final Mass, Velocity Dispersion, Effective Radius, Central
    Surface Brightness, Ellipticity, Position Angle. The final model
    is marked with a star.} 
  \label{tab:test}
  \begin{tabular}{l c c c c c c c c}
    \hline
    Model & $R_{\rm pl}$ & $M_{\rm pl}$ & $M_{V}$ & $\sigma_{\rm
      los}$  & $R_{\rm h}$  & $\mu_0$ & $\epsilon$ & $PA$ \\  
    & [pc] & $[\times 10^7\,\mathrm{M}_\odot]$&$[\times
    10^5\,\mathrm{M}_\odot]$ & [km s$^{-1}$]& [pc] & [mag] &
    & [degree] \\ 
    \hline
    1 & 588 & 1.86 & 2.46 & 6.65 & 535.3 & 26.85  & 0.445 & 74.4\\
    2 & 653 & 2.38 & 2.14 & 8.14 & 555.2 & 27.04  & 0.266 & 59.4\\
    3 & 657 & 2.63 & 2.86 & 7.21 & 545.8 & 26.687 & 0.257 & 96.9\\
    4 & 712 & 3.25 & 2.79 & 8.13 & 554.1 & 26.766 & 0.292 & 70.8\\
    5 (*) & 653 & 2.47 & 2.45 & 7.58 & 545.7 & 26.847 & 0.466 & 71.8\\
    \hline
  \end{tabular}
\end{table*}

\begin{figure} 
  \centering
  \includegraphics[width=6cm]{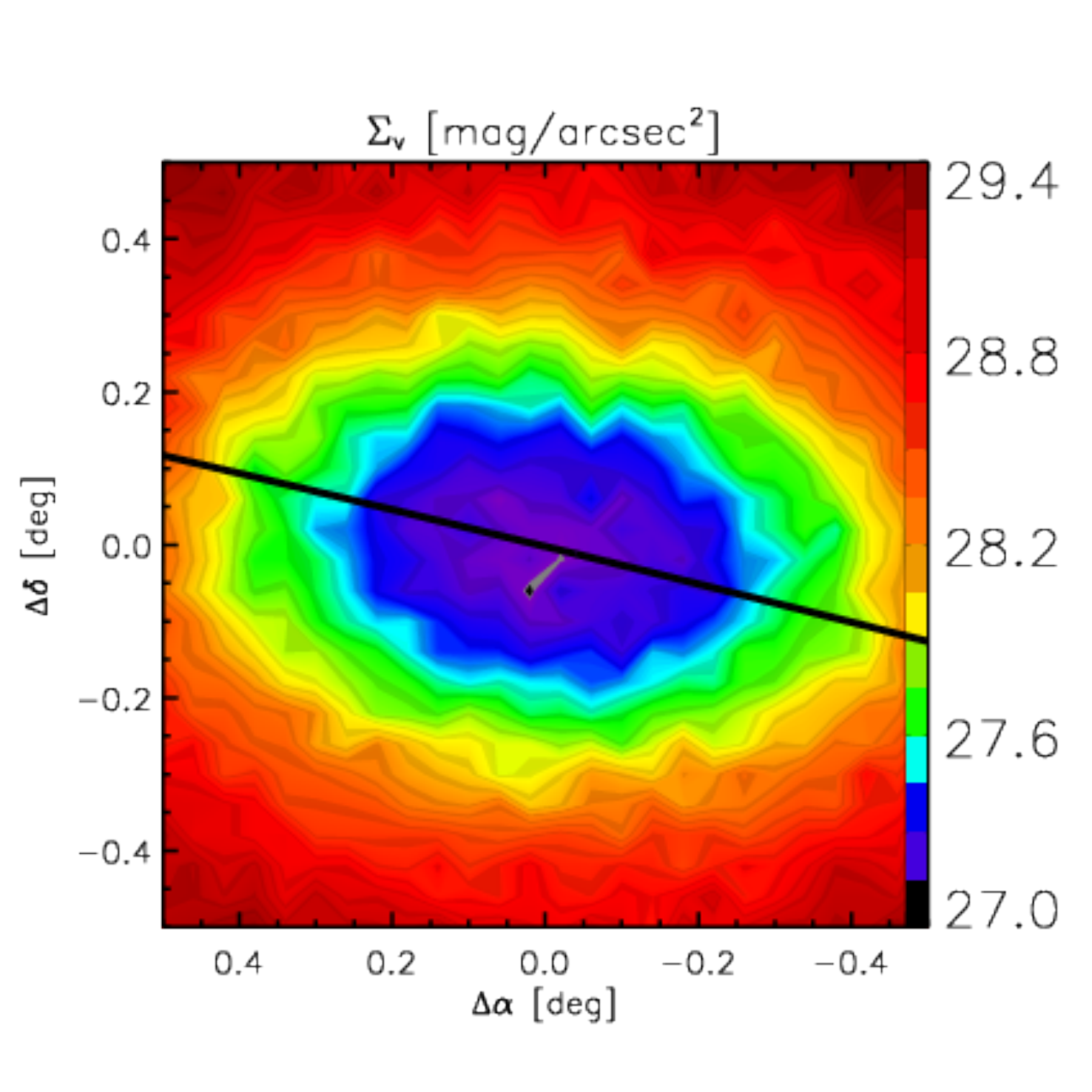}
  \includegraphics[width=6cm]{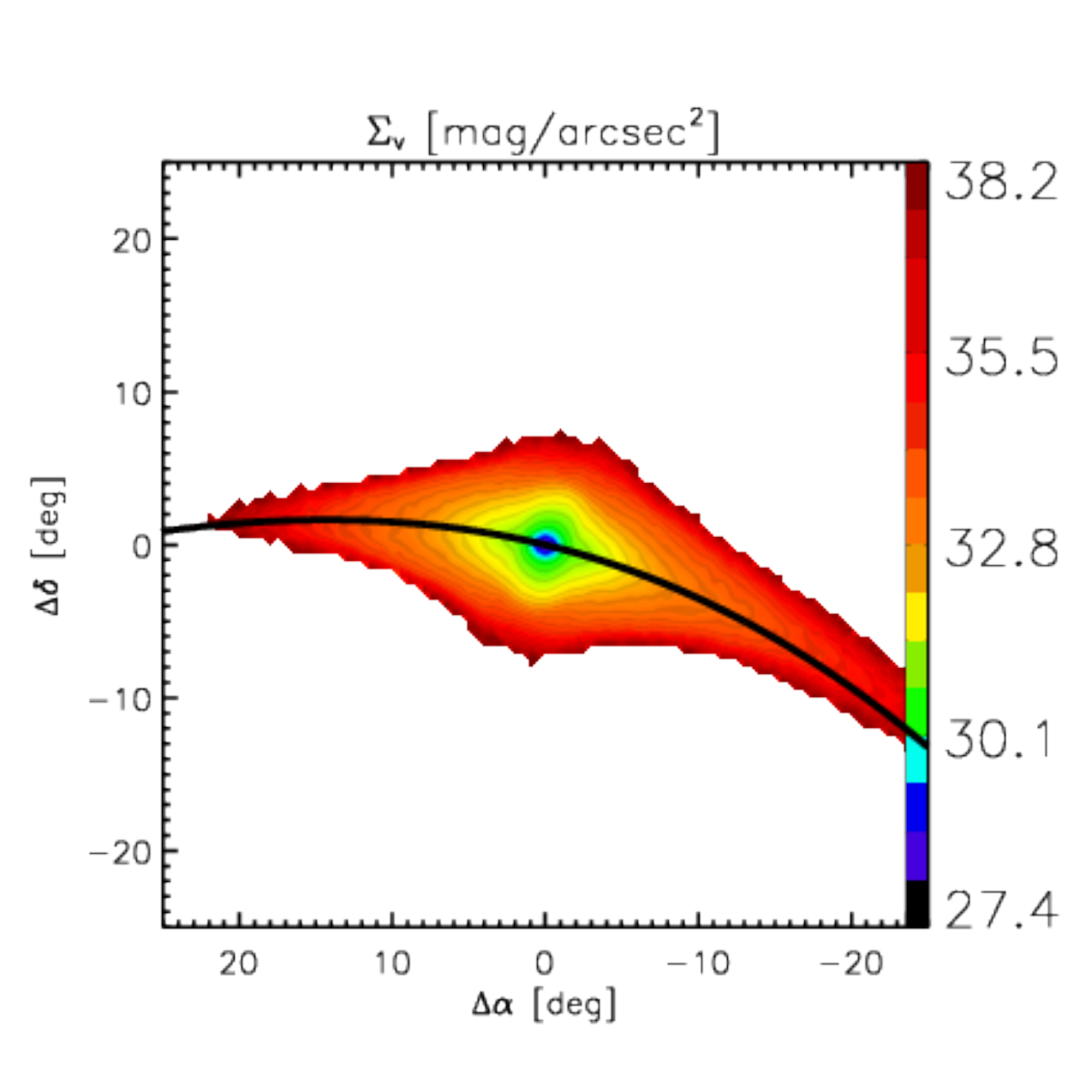}
  \caption{2D contour plot of the surface brightness of the
    best-matching model using a generic M/L of unity.  Top
    panel: The innermost square-degree shown with a resolution of 144
    arcsec per pixel.  Compare with Fig.~\ref{fig:proy_orb}.  Bottom
    panel: 50x50 degrees showing the extend of the tidal tails along
    the orbit.  The resolution is 0.5 degree.  It is important to note
    that yellow to red contours in this panel have surface
    brightnesses below any normal observational detection limit.}  
    \label{fig:2Dcont}
\end{figure}

We compare the final values of mass, surface brightness, effective
radius, and velocity dispersion, plus ellipticity and position angle,
of the candidates with the observed values to choose the final model.
From these, the fifth model, with the pair of initial parameters
$R_{\rm pl} = 653$ pc and $M_{\rm pl} = 2.47
\times\,10^7\,\mathrm{M}_\odot$, produces an object that matches the
observed properties of CVn~I dSph at the end of the simulation, with
an infall time of $10$~Gyr, with values within the observational
errors: a final mass of $2.45 \times 10^5\,\mathrm{M}_\odot$, central
surface brightness of $25.7$~mag\;arcsec$^{-2}$, effective radius of
$553.8$~pc, and a velocity dispersion with the value
$7.61$~km\;s$^{-1}$ (see Fig.~\ref{fig:hist_disp} and compare with
figure~8 of \citet{simon_2007}).  

A summary of the properties of the remnant can be found in
Tab.~\ref{tab:test} and a 2D contour plot of the object in
Fig.~\ref{fig:2Dcont}.  In the top panel we see that our final
object has the correct elongation along the orbit, that is position
angle, as well as the correct ellipticity.  In the lower panel we
increase the field of view (and lower the resolution as well) to
show how the tidal tails align with the chosen orbit and how
compressed they appear because the object is close to the apo-centre
of its orbit.  We point out that the colour scale is different in the
top and lower panel.  Furthermore, it quite clear that the far away
tails shown in the lower panel have surface brightnesses far below
any observational detection limit available today.

\section{Conclusions}
\label{sec:conc}

There are different scenarios that explain the formation and evolution
of dSphs like CVn~I.  Some use dwarf galaxies embedded in dark halos
\citep{mayer_2007, d'onghia_2009} or dissolving star clusters inside
said dark haloes \citep{assmann_2013a, assmann_2013b, alarcon_2018}. 
Unlike them, and in the same vein as \citet{blana_2015} and
\citet{domingez_2016}, in this work we offer an alternative way to
reproduce the characteristics of CVn~I, without using DM nor any
modification to the Newtonian gravity. 

In the scenario we present here, we use a DM-free progenitor at
the start of our simulations and explain the measured elongation of
the dwarf satellite as tidal distorsions due to the disruptive forces
acting on the object, that is peri-centre passages and/or disc
passages.  The high velocity dispersion is not caused by a massive
DM halo but is the effect of many unbound stars surrounding the
object mimicking a Gaussian-like distribution of velocities similar
to an object in equilibrium.  The actual initial object in this
scenario could either be a DM-free object from the start like an
extended star cluster or a tidal dwarf galaxy, or it could be a former
DM dominated object which has lost all or most of its dark matter due
to the tidal forces of the MW.  If there were other mechanisms that
played a r\^{o}le in the evolution of CVn~I (e.g. close encounters
with other satellites, gas loss via reionization or ram pressure
stripping), then the initial mass that we present here would be a
lower limit for the real initial mass of this dSph.
 
As we use phase-space particles instead of stars and completely
neglect the contribution of any gas component in the early stages of
CVn~I's history, it might be interesting for future projects to study
the impact of gas stripping or stellar feedback as contributors to its
evolution, mass loss, and subsequent tidal disruption by the MW.  On
the other hand we use a MW potential as it is seen now throughout the
simulation.  In reality we expect that the mass of the MW was growing
during the last $10$~Gyr to reach its value now.  Taking this effect
into account would help the survival of the initial object.
Furthermore, we do neglect dynamical friction.  This would place the
initial object on a more distant orbit to begin with, which again
would help in the survival of the satellite.

We were able to find a pair of initial parameters of a Plummer sphere
that reproduces the current observational properties of CVn~I dSph
(Tab.~\ref{tab:test}).  With a initial mass of $2.47 \times 10^7\,
\mathrm{M}_\odot$ and Plummer radius of $653$~pc, we obtain a remnant
with a final mass of $2.45 \times 10^5\,\mathrm{M}_\odot$, central
surface brightness of $26.9$~mag\;arcsec$^{-2}$, an effective radius
of $545.7$~pc, and a velocity dispersion with the value
$7.58$km\;s$^{-1}$, properties that matches the observational
parameters within the measurement errors.

To achieve this, first we had to find an orbit for this dwarf galaxy. 
Under the assumption that Canes Venatici~I is a DM-free object, its
elongation, as observed by \citet{okamoto_2012}, would be a
consequence of the tidal disruption that this galaxy is suffering
under the influence of the MW's gravitational potential and,
therefore, this dwarf's major axis would be aligned with it's orbital
path.  If that weren't the case, we could choose any other orbit, as
long as the resulting object has the same radial velocity as it is
observed in CVn~I.  Its high velocity dispersion can be explained by
tidal debris surrounding the object, so the chosen orbit has to
produce a high mass loss, and as long as the object is close to the
apo-centre, the velocity dispersion will be boosted. 

With these three points in mind, we choose a very eccentric orbit
($\epsilon \sim 0.85$), with a small perigalactic distance
($D_{\rm p} \sim 18.8$~kpc) and an apogalactic distance close to the
current distance of CVn~I dSph ($D_{\rm a} \sim 243$~kpc,
$D_{\rm CVn I} \sim 224$~kpc).  Using this orbit, we explored the
space of initial parameters for the Plummer sphere, covering from
{$320$~pc $\le R_{\rm pl} \le 890$~pc} to $4 \times
10^6\,\mathrm{M}_\odot \le M_{\rm pl} \le 10^8\, \mathrm{M}_\odot $. 
With these results we found the power laws that represent the pairs of
initial parameters that reproduce the respective observable. 
In the region where all the power laws come close to each other
($600\,\mathrm{pc} \le R_{\rm pl} \le 700\,\mathrm{pc}$,
$20.5\times10^6\,\mathrm{M_\odot} \le M_{\rm pl} \le 30\times
10^6\,\mathrm{M_\odot}$), we found the final model described above,
the one that manages to reproduce the current observational parameters
of CVn~I dSph. 

This study shows, that it is possible to find a DM free progenitor,
which matches all the structural and dynamical observables of Canes
Venaticii~I.  This finding, by no means, claims to be the only
possible scenario to explain this particular dwarf galaxy, but simply
adds an additional possible formation and/or evolution channel,
without invoking alterations to the gravity law.

\section*{Acknowledgements}

DRMC, MF, AGAJ, CAA and FUZ acknowledge financial help through
Fondecyt regular No.~1180291 and BASAL 'Centro de Astrofisica y
Tecnologias Afines' (CATA) No. AFB-170002. 
MF additionally acknowledges financial support through  Conicyt
PII20150171 and Quimal 170001. 
AGAJ acknowledges financial support from Carnegie Institution of
Washington with its Carnegie-Chile Fellowship. 

\bibliographystyle{aa}
\bibliography{cvp-05}

\begin{appendix}

\section{Search method for the orbits}
\label{sec:ap1}

The search method goes as follows:
\begin{enumerate}
\item The search area is defined  as the zone in parameter space with
  proper motions with $(\mu_{ \delta,c}-R) \leq \mu_\delta \leq
  (\mu_{\delta,c}+R)$ and $(\mu_{\alpha,c}-R) \leq \mu_\alpha \leq
  (\mu_{\alpha,c}+R)$, where R is the search radius. 
\item The search area is divided in $N^2$, equally spaced, pairs of
  proper motions $(\mu_{\alpha,i},\mu_{\delta,j})$, where: 
  \begin{eqnarray}
    \mu_{\alpha,i} = (\mu_{\alpha,c}-R)+i \cdot \frac{2R}{N-1}\,,i = 0,1,..,N-1,\\
    \mu_{\delta,j} = (\mu_{\delta,c}-R)+j \cdot \frac{2R}{N-1}\,,j = 0,1,..,N-1.
  \end{eqnarray}

\item For each of these pairs of proper motions, the orbit is
  calculated using a simple point mass integrator. 
\item This orbit is projected on the sky and its position angle
  measured at the current coordinates of CVn~I.  
\item The difference $\Delta PA$ between CVn~I's position angle and
  the projected orbit's inclination (see Fig.~\ref{fig:proy_orb}) is
  calculated \label{item}.   
\item The N pairs $(\mu_{\alpha,i},\mu_{\delta,i})$ that gave values
  with  $ 0 \leq\Delta PA < 5$  are saved (since these are the orbits
  that closely match the position angle of CVn~I dSph)  and the rest
  are discarded. 
\item For each of these $N$ pairs $(\mu_{\alpha,i},\mu_{\delta,i})$,
  this process is repeated, now with: 
  \begin{eqnarray}
    \nonumber
    \mu_{\alpha,c} =& \mu_{\alpha,i},\\
    \nonumber
    \mu_{\delta,c} = &\mu_{\delta,i},\\
    \nonumber
    R =& \frac{R}{N-1}.
  \end{eqnarray}
\end{enumerate}

To find the candidates using this method, the initial values chosen
were: 
\begin{eqnarray}
  \nonumber
  \mu_{\alpha,c} = &0, \\
  \nonumber
  \mu_{\delta,c} = &0, \\
  \nonumber
  R = &0.2, \\
  \nonumber
  N =&7.\ 
\end{eqnarray}
    
The value of N can be increased if we needed more resolution or faster
results, but $N=7$ is a good compromise, so at the end the 343 orbits
showed us the region on parameter space where the orbit has the same
inclination as the observed object. 

The technique described is flexible enough to use other properties of
the orbit to select regions of this space, like perigalactic or
apogalactic distance. Since CVn~I dSph has a high velocity dispersion,
the chosen orbit should be the one that will give the highest boost on
this value of the final object. 

\end{appendix}

\end{document}